\documentclass[
 aps,pra,
 twocolumn,
 superscriptaddress,
 longbibliography,
 nofootinbib
]{revtex4-2}

\usepackage{amsmath,amssymb,amsfonts,amsthm}
\usepackage{graphicx}
\graphicspath{{figures/}}
\usepackage{quantikz}
\usepackage{xcolor}
\usepackage{hyperref}
\hypersetup{colorlinks=true,linkcolor=blue,citecolor=blue,urlcolor=blue}
\usepackage{orcidlink}

\newcommand{\Tr}{\mathrm{Tr}}
\providecommand{\ket}[1]{\lvert #1 \rangle}
\providecommand{\bra}[1]{\langle #1 \rvert}
\newcommand{\AD}{\mathrm{AD}}
\newtheorem{proposition}{Proposition}
\newtheorem{corollary}{Corollary}
\newcommand{\authornote}{Contact:
\href{mailto:ltkhanh@hcmus.edu.vn}{ltkhanh@hcmus.edu.vn}}

\begin{document}

\title{Zero- Versus Infinite-Temperature Damping in Variational Quantum
Circuits: Feature Scale, Sampling Cost, and Frame Gauge}

\author{Vu-Quoc-Minh Nguyen\,\orcidlink{0009-0007-2748-5107}}
\affiliation{Faculty of Electronics and Telecommunications, University of Science,
Vietnam National University Ho Chi Minh City (VNU-HCM), Ho Chi Minh City 70000,
Vietnam}

\author{Tuan-Vu Truong\,\orcidlink{0009-0005-6357-6060}}
\affiliation{Identity Quantum Computing JSC, 117 Tran Duy Hung Street, Hanoi 10000,
Vietnam}

\author{Hoang-Long Nguyen\,\orcidlink{0009-0005-7922-9266}}
\affiliation{Identity Quantum Computing JSC, 117 Tran Duy Hung Street, Hanoi 10000,
Vietnam}

\author{Trung-Khanh Le\,\orcidlink{0000-0003-0935-6743}}
\thanks{\authornote}
\affiliation{Faculty of Electronics and Telecommunications, University of Science,
Vietnam National University Ho Chi Minh City (VNU-HCM), Ho Chi Minh City 70000,
Vietnam}

\date{\today}

\begin{abstract}
The Pauli twirl of amplitude damping (AD) is generalized amplitude damping at
infinite temperature: it keeps the contraction of AD and removes its
non-unital term, so comparing the two in variational circuits isolates the
zero-temperature bias, which acts mainly through the scale of the features.
For random parameters, features under AD settle on a floor, which at strong
damping is set by the last layer, has a closed form, and at fixed $T_1$ falls
with temperature as $\tanh(\hbar\omega/2k_BT)$; under the twirls they shrink
by a constant factor per layer, up to eight qubits. A trainable output scale
removes most of the resulting accuracy differences, leaving AD ahead of its
twirls by at most about three percentage points in our simulations; what it
removes reappears as a cost in measurement shots: trained and tested with
$10^3$ shots per image, a four-qubit classifier under AD at $p=0.3$ stays
within 1.5 points of noiseless accuracy, while the twirled classifiers lose up
to 33. At weaker damping the separation depth grows roughly as
$(np)^{-1}\ln(1/p)$. The damping direction is a gauge when the damping follows
complete entangling layers and the circuit boundaries are trainable; in an
eigensolver it becomes physical inside a decomposed two-qubit gate.
\end{abstract}

\maketitle

\section{Introduction}
\label{sec:intro}

Variational quantum algorithms and quantum machine learning models are built
from parameterized quantum circuits and are expected to run on noisy
intermediate-scale quantum devices~\cite{preskill2018nisq,
cerezo2021variational,bharti2022noisy}. Whether such models can be trained
depends on how the output concentrates. Gradients of deep random circuits,
and of global cost functions even in shallow ones, vanish exponentially with
the number of qubits~\cite{mcclean2018barren,cerezo2021cost,
larocca2025barren}. Unital noise adds a concentration with
depth: the state approaches the maximally mixed state, and local expectation
values and their gradients vanish exponentially~\cite{wang2021noiseinduced,
stilckfranca2021limitations}. Non-unital noise behaves differently. Amplitude
damping (AD), which models energy relaxation ($T_1$
decay)~\cite{krantz2019quantum}, a dominant source of noise in
superconducting qubits~\cite{quek2024exponentially}, drives the state toward a
pure state. Under non-unital noise, cost functions of local observables do not
exhibit barren plateaus~\cite{mele2026noise}, and the output distribution of
random circuits never anticoncentrates~\cite{fefferman2024effect}. For
general noise, the gradients with respect to early layers still
vanish~\cite{schumann2024emergence}, and amplitude damping, a
Hilbert--Schmidt-contractive map, need not produce noise-induced barren
plateaus, although its cost concentrates onto a noise-induced limit set beyond
logarithmic depth~\cite{singkanipa2025beyond}. On hardware dominated by
$T_1$ decay, gradient magnitudes saturate instead of decaying
exponentially~\cite{schmitt2026experimental}, and
engineered dissipation can restore trainability~\cite{sannia2024engineered}.

Noise tailoring removes this structure. Randomized compiling, or Pauli
twirling, turns an arbitrary gate error into a stochastic Pauli
channel~\cite{wallman2016noise,hashim2021randomized}, which is easier to
characterize and mitigate~\cite{temme2017error,cai2023quantum}, and a
Clifford twirl turns it into depolarizing noise~\cite{dankert2009exact}. The
Pauli twirl of AD keeps its contraction and discards its non-unital term: it
is the equal mixture of damping toward $\ket0$ and toward $\ket1$, that is,
generalized amplitude damping at infinite temperature. Comparing AD with its
Pauli twirl in a variational model therefore asks what the zero-temperature
bias of the damping does. Van~Rossum \textit{et al.}~\cite{vanrossum2026biased}
made this comparison in data re-uploading circuits and in a variational
eigensolver. They found that the twirled channels reduce the range of outputs
the model can fit, which they use as a measure of expressivity, reduce
gradient magnitudes, and degrade eigensolver solutions, and that in the
eigensolver damping in the reversed direction performs significantly worse
than ordinary damping. They concluded that ``commonly used noise reshaping
techniques, such as Pauli twirling, may inadvertently degrade the performance
of variational algorithms''~\cite{vanrossum2026biased}.

Here, we show that the bias acts on variational models mainly through the
scale of their features, and that its direction is a gauge when the damping
acts after complete entangling layers and trainable gates and the readout
absorb the frame at the boundaries of the circuit. For random parameters, local
features under AD settle on a floor, which at strong damping is set by the
last layer and has a closed form, whereas under the twirls they shrink by a
constant factor per layer; we establish this at widths up to eight qubits. A
trainable affine readout absorbs a contraction, so it removes most of the
differences in fit and accuracy in exact
simulation~\cite{wang2024errormitigation,vovrosh2021simple}. What it removes
reappears as a cost in measurement shots, which we measure: trained and tested with $10^3$
shots per image, the classifier under AD stays close to the noiseless
accuracy, whereas the twirled classifiers lose up to 33 percentage points.
The separation needs a depth that grows roughly as $(np)^{-1}\ln(1/p)$ for
$n$ qubits at damping strength $p$, so at the damping of current hardware it
is a property of very deep circuits. At finite temperature and fixed $T_1$ the floor falls as
$\tanh(\hbar\omega/2k_BT)$, and the Pauli twirl is its infinite-temperature
end.
The direction of the damping, in turn, is a label that a change of frame can
move. We give conditions under which reversing it is a reparametrization of
the model, implemented by a Pauli frame or by spin time reversal. The frame
propagation is that of the parameter symmetries of
Ref.~\cite{fontana2022nontrivial}; here it relates two noise models. Time
reversal leaves every single-qubit rotation invariant, so the encoding never
protects the direction. Only the fixed multi-qubit gates, the placement of the
noise relative to them, and the boundaries of the circuit can make the
direction physical. Where the frame is absorbed, the initialization is
symmetric, and the optimizer is equivariant, trained models under the two
directions are identically distributed.

We test these statements on three models: the single-qubit re-uploading fits
of Ref.~\cite{vanrossum2026biased}, a four-qubit re-uploading classifier on
MNIST and Fashion-MNIST, and the three-qubit eigensolver of
Ref.~\cite{vanrossum2026biased}. With a trainable output scale, AD and its
twirls fit equally well and classify to within about one percentage point at
four layers of strong damping. The direction effect in the eigensolver exists
because the damping acts inside the decomposed two-qubit gate, it vanishes
exactly when the same damping follows the gate, and it persists when the
transverse field dominates; on an open chain the sign of the coupling is itself
a gauge of the ansatz, so the effect cannot depend on it. Beyond the scale, the
bias leaves AD ahead of its twirl by up to three percentage points in exact
simulation; at $p=0.1$ this margin grows from 0.9 points at four layers to 2.6
at the separation depth. At three and four layers it also gives the
classifier a small gain over the noiseless circuit in exact simulation, which
does not survive training and testing with $10^3$ shots per image, and on
MNIST it keeps the input sensitivity from growing with depth; on
Fashion-MNIST it does not.

The paper is organized as follows. Section~\ref{sec:setting} introduces the
channels and the three models. Section~\ref{sec:scale} treats the output scale
and Sec.~\ref{sec:gauge} the gauge. Section~\ref{sec:cost} shows how the
non-unital term sets the feature scale and the sampling cost, and what it
changes beyond them, and Sec.~\ref{sec:conclusion} concludes.

\section{Channels and models}
\label{sec:setting}

\subsection{Channels}
\label{sec:channels}

A single-qubit channel with Kraus operators $K_k$,
\begin{equation}
\mathcal N(\rho)=\sum_k K_k\,\rho\,K_k^\dagger ,
\label{eq:kraus}
\end{equation}
acts on the Bloch vector as an affine map~\cite{nielsen2010quantum},
\begin{align}
\mathbf r\;&\mapsto\;T\mathbf r+\mathbf t,
\label{eq:bloch}\\
T_{ij}&=\tfrac12\Tr\bigl[P_i\,\mathcal N(P_j)\bigr],\qquad
t_i=\tfrac12\Tr\bigl[P_i\,\mathcal N(I)\bigr],
\label{eq:bloch_Tt}
\end{align}
where $P_i\in\{X,Y,Z\}$. The channel is unital, $\mathcal N(I)=I$, if and only
if $\mathbf t=\mathbf 0$. We compare the conditions of
Table~\ref{tab:channels}, as in Ref.~\cite{vanrossum2026biased}. Amplitude
damping of strength $p$ has
\begin{equation}
T_{\AD}=\mathrm{diag}(c,\,c,\,c^2),\qquad
\mathbf t_{\AD}=(0,\,0,\,p),
\label{eq:ad}
\end{equation}
with $c=\sqrt{1-p}$.
Its Pauli twirl~\cite{dur2005standard,wallman2016noise} has the same
contraction and no non-unital term,
\begin{equation}
T_{\rm Pauli}=T_{\AD},\qquad \mathbf t_{\rm Pauli}=\mathbf 0 ,
\label{eq:pauli}
\end{equation}
and its Clifford twirl is the depolarizing channel~\cite{dankert2009exact},
\begin{equation}
T_{\rm Depol}=\lambda_d\,\mathbb I_3,\qquad \mathbf t_{\rm Depol}=\mathbf 0,
\label{eq:depol}
\end{equation}
with $\lambda_d=\frac13(2c+c^2)$, the mean of the contraction factors of AD.
The reversed channel damps toward $\ket1$,
\begin{equation}
\AD^{\rm flip}=\mathcal X\circ\AD\circ\mathcal X,\qquad
\mathcal X(\rho)=X\rho X ,
\label{eq:adflip}
\end{equation}
with the same $T$ and $\mathbf t=(0,0,-p)$. Since $Z\AD Z=\AD$ and
$X\AD X=Y\AD Y=\AD^{\rm flip}$, the Pauli twirl is the equal mixture
\begin{equation}
\AD_{\rm Pauli}=\tfrac12\bigl(\AD+\AD^{\rm flip}\bigr),
\label{eq:twirl_mix}
\end{equation}
which is generalized amplitude damping at infinite
temperature~\cite{nielsen2010quantum}. Comparing AD with its Pauli twirl
therefore compares damping at zero and at infinite temperature. More
generally, generalized amplitude damping with excited-state population $N$ has
the same $T$ and $\mathbf t=(0,0,p(1-2N))$; the frame $\mathcal X$ maps $N$ to
$1-N$, and AD, its Pauli twirl, and AD$^{\rm flip}$ are the cases $N=0$,
$\frac12$, and 1. As in
Ref.~\cite{vanrossum2026biased}, the twirled conditions replace the channel by
its twirl at every insertion; no twirling protocol implements this after
non-Clifford gates, so the twirled channels serve as model channels with the
contraction of AD. In simulations of error correction, this Pauli twirling
approximation is accurate for some stabilizer circuits~\cite{geller2013efficient}
but misses features of amplitude damping that other efficiently simulable
channels capture~\cite{gutierrez2013approximation}.

\begin{table*}[t]
\caption{Noise conditions at strength $p$, with $c=\sqrt{1-p}$ and
$\Lambda=1-\frac13(2c+c^2)$. Unital channels are given by their Pauli
weights, with $p_0=1-\sum_{k\neq0}p_k$.}
\label{tab:channels}
\begin{ruledtabular}
\begin{tabular}{lll}
channel & Kraus operators & $(T,\mathbf t)$\\
\colrule
None & $K_0=I$ & $\mathbb I_3,\;\mathbf 0$\\
AD & $\mathrm{diag}(1,c)$, $\sqrt p\,\ket0\bra1$ &
 $\mathrm{diag}(c,c,c^{2}),\;(0,0,p)$\\
AD$^{\rm flip}$ & $\mathrm{diag}(c,1)$, $\sqrt p\,\ket1\bra0$ &
 $\mathrm{diag}(c,c,c^{2}),\;(0,0,-p)$\\
Pauli & $p_x=p_y=\frac p4$, $p_z=\frac14(2-p-2c)$ &
 $\mathrm{diag}(c,c,c^{2}),\;\mathbf 0$\\
Depol & $p_x=p_y=p_z=\frac\Lambda4$ & $\frac13(2c+c^2)\,\mathbb I_3,\;\mathbf 0$\\
\end{tabular}
\end{ruledtabular}
\end{table*}

\subsection{Single-qubit re-uploading fits}
\label{sec:model_vr}

Our first model reproduces the setting of Fig.~2(a) of
Ref.~\cite{vanrossum2026biased}. A single qubit starts in $\ket0$ and passes
through $L=2$ layers. With $\mathcal W_l=\mathcal U(R_ZR_YR_Z)$ the trainable
blocks, $\mathcal S_x=\mathcal U(R_X(x))$ the encoding, and $\mathcal U(V)\colon
\rho\mapsto V\rho V^\dagger$, the channel acts after every block,
\begin{equation}
\Phi_{x,\theta}=\mathcal N\circ\mathcal W_2\circ\mathcal N\circ\mathcal S_x
\circ\mathcal N\circ\mathcal W_1\circ\mathcal N\circ\mathcal S_x\circ\mathcal N
\circ\mathcal W_0 ,
\label{eq:vr_map}
\end{equation}
with five insertions and nine parameters. The output of the raw model is
\begin{equation}
f(x)=\Tr\bigl[Z\,\Phi_{x,\theta}(\ket0\bra0)\bigr].
\label{eq:vr_raw}
\end{equation}
As targets we use degree-2 trigonometric polynomials,
\begin{equation}
g'(x)=a_1\cos x+b_1\sin x+a_2\cos2x+b_2\sin2x ,
\label{eq:vr_target}
\end{equation}
with $a_k,b_k$ drawn uniformly from $[-1,1]$, normalized to unit range and zero
mid-range and then scaled to a range $R$, and sampled at 250 points in
$[-2\pi,2\pi]$. Ref.~\cite{vanrossum2026biased} measures expressivity as the
largest $R$ for which Adam (learning rate 0.20, $\beta_1=0.45$) reaches a mean
squared error below $10^{-5}$ within 150 steps. We use 200 seeded targets, the
same for every channel, and compare the raw model with a rescaled model,
\begin{equation}
f_{s,b}(x)=s\,f(x)+b ,
\label{eq:vr_scaled}
\end{equation}
in which $s$ and $b$ are trained together with the circuit. The details of
our reading of the protocol are given in Appendix~\ref{app:vr}.

\subsection{Four-qubit re-uploading classifier}
\label{sec:model_clf}

\begin{figure*}[!t]
\centering
\begin{quantikz}[column sep=3.2pt, row sep=10pt, thin lines]
\lstick{$\ket{0}$}
 & \gate{R_Z}\gategroup[4,steps=10,style={dashed,rounded corners,
 inner xsep=3pt,inner ysep=3pt},background,
 label style={label position=below,anchor=north,yshift=-0.35cm}]
 {\small one layer, repeated $\times L$}
 & \gate{R_Y} & \gate{R_Z} & \ctrl{1} & & & \targ{}
 & \gate[4]{\mathcal N^{\otimes n}} & \gate{R_X(a_1)}
 & \gate[4]{\mathcal N^{\otimes n}}
 & \gate{R_Z} & \gate{R_Y} & \gate{R_Z}
 & \gate[4]{\mathcal N^{\otimes n}} & \meter{} & \rstick{$z_1$} \\
\lstick{$\ket{0}$}
 & \gate{R_Z} & \gate{R_Y} & \gate{R_Z} & \targ{} & \ctrl{1} & &
 & & \gate{R_X(a_2)} & & \gate{R_Z} & \gate{R_Y} & \gate{R_Z} & &
 \meter{} & \rstick{$z_2$} \\
\lstick{$\ket{0}$}
 & \gate{R_Z} & \gate{R_Y} & \gate{R_Z} & & \targ{} & \ctrl{1} &
 & & \gate{R_X(a_3)} & & \gate{R_Z} & \gate{R_Y} & \gate{R_Z} & &
 \meter{} & \rstick{$z_3$} \\
\lstick{$\ket{0}$}
 & \gate{R_Z} & \gate{R_Y} & \gate{R_Z} & & & \targ{} & \ctrl{-3}
 & & \gate{R_X(a_4)} & & \gate{R_Z} & \gate{R_Y} & \gate{R_Z} & &
 \meter{} & \rstick{$z_4$}
\end{quantikz}
\caption{The classifier at $n=4$, drawn for one layer and the final block.
Each $R_ZR_YR_Z$ triple carries three trainable angles; the encoding column
re-uploads the angle vector $\mathbf a$ in each of the $L$ layers. The channel
acts on every qubit after the entangler, after the encoding, and after the
final block, giving $2L+1$ insertions per qubit. The features
$z_j=\langle Z_j\rangle$ enter a trainable linear readout.}
\label{fig:circuit}
\end{figure*}
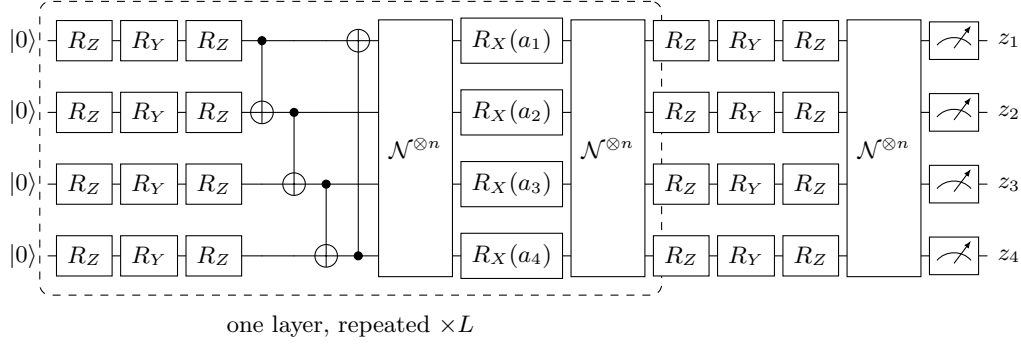

Our second model is a data re-uploading
classifier~\cite{perezsalinas2020data,schuld2021encoding} on $n=4$ qubits
(Fig.~\ref{fig:circuit}). We standardize each MNIST image~\cite{lecun1998gradient} of the digits
$\{1,3,5\}$, reduce it to four principal components $u_j$, and map each
component to an angle by a min--max transformation,
\begin{equation}
a_j=2\pi\,\frac{u_j-u_j^{\min}}{u_j^{\max}-u_j^{\min}} .
\label{eq:angles}
\end{equation}
The standardization, the principal components, and the extrema are all
fitted on 1000 training images and applied unchanged to all 3037 test images
of the three digits; 14 of the 12\,148 test angles fall outside $[0,2\pi]$ and
are clipped. With the trainable blocks, the encoding, and the ring
entangler,
\begin{align}
\mathcal R^{(\ell)}&=\bigotimes_{j=1}^{n}R_Z R_Y R_Z,\qquad
S(\mathbf a)=\bigotimes_{j=1}^{n}R_X(a_j),
\label{eq:blocks}\\
E&=\prod_{j=1}^{n}\mathrm{CNOT}_{j,\,j+1},
\label{eq:ring}
\end{align}
with the qubit index taken modulo $n$, a layer performs
\begin{equation}
\mathcal L^{(\ell)}_{\mathbf a}=\mathcal N^{\otimes n}\circ
\mathcal U\bigl(S(\mathbf a)\bigr)\circ\mathcal N^{\otimes n}\circ
\mathcal U\bigl(E\,\mathcal R^{(\ell)}\bigr),
\label{eq:layer}
\end{equation}
and the circuit is
\begin{equation}
\Phi_{\mathbf a,\theta}=\mathcal N^{\otimes n}\circ\mathcal U\bigl(\mathcal R^{(L)}\bigr)
\circ\mathcal L^{(L-1)}_{\mathbf a}\circ\cdots\circ\mathcal L^{(0)}_{\mathbf a},
\label{eq:circuit}
\end{equation}
with $2L+1$ insertions of the channel per qubit. The features are the
single-qubit expectation values
\begin{equation}
z_j(\mathbf a,\theta)=\Tr\bigl[Z_j\,\Phi_{\mathbf a,\theta}
\bigl(\ket0\bra0^{\otimes n}\bigr)\bigr],
\label{eq:features}
\end{equation}
and the scores of the three classes are
\begin{equation}
\ell(\mathbf a)=
\begin{cases}
W\mathbf z+\mathbf b & \text{(raw)},\\
W\,\mathrm{diag}(\boldsymbol\sigma)^{-1}(\mathbf z-\boldsymbol\mu)+\mathbf b
 & \text{(standardized)},
\end{cases}
\label{eq:readout}
\end{equation}
where $\boldsymbol\mu$ and $\boldsymbol\sigma$ are the mean and the population
standard deviation of the features over the $M=1000$ training images,
\begin{align}
\mu_j&=\frac1M\sum_{m=1}^{M}z_j(\mathbf a_m,\theta),
\label{eq:mu}\\
\sigma_j^2&=\frac1M\sum_{m=1}^{M}\bigl[z_j(\mathbf a_m,\theta)-\mu_j\bigr]^2 .
\label{eq:musigma}
\end{align}
They are
recomputed at every step, with gradients flowing through them, and the values
at the final parameters are applied to the test images. The standardized
readout has no parameter beyond those of the raw one; it only fixes the scale
at which $W$ acts. No regularizing constant is added to $\boldsymbol\sigma$;
in one run a feature's spread fell to $5\times10^{-5}$ and the training
diverged (Table~\ref{tab:clf}). Training minimizes the softmax cross-entropy with
Adam~\cite{kingma2014adam} at learning rate $0.05$ for 100 full-batch steps
unless stated otherwise, without weight decay. The circuit angles start
uniformly in $[0,2\pi)$, the readout weights from $0.5\,\mathcal N(0,1)$, and
the bias at zero. We use $p=0.3$ and simulate the density matrix
exactly in PyTorch~\cite{paszke2019pytorch}, so all expectation values are free of sampling error. As a second
data set we use the Fashion-MNIST~\cite{xiao2017fashion} classes T-shirt, pullover, and coat, with
3000 test images. Three classes and 1000 training images keep exact
density-matrix training of eight initializations per condition affordable. For
reference, logistic regression, a $k$-nearest-neighbor
classifier ($k=15$), and a support-vector machine with a radial kernel, trained
on the same angles, reach 0.856, 0.879, and 0.886 on MNIST and 0.733, 0.758,
and 0.748 on Fashion-MNIST. We quantify the input sensitivity of a trained
classifier by
\begin{equation}
\chi=\bigl\langle\lVert\nabla_{\mathbf a}\ln q_y(\mathbf a)\rVert\bigr\rangle,
\label{eq:sens}
\end{equation}
where $q_y$ is the softmax probability of the true class $y$ and the average
runs over the test images.

For the finite-shot runs, each image is measured $N_s$ times in the
computational basis, and $\hat z_j$ is the mean of the $N_s$ outcomes $\pm1$ of
qubit $j$. At test time we sample the outcomes from the exact distribution of
the measured state, five times per model, and in every draw we re-estimate
$\boldsymbol\mu$ and $\boldsymbol\sigma$ from $N_s$ shots of each training
image, as one would on hardware. In training we replace each feature by
\begin{equation}
\hat z_j=z_j+\sqrt{\frac{1-z_j^2}{N_s}}\;\varepsilon_j,\qquad
\varepsilon_j\sim\mathcal N(0,1),
\label{eq:shot_train}
\end{equation}
redrawn at every step and for every image, and compute $\boldsymbol\mu$ and
$\boldsymbol\sigma$ from the noisy features; gradients flow through $z_j$. This
Gaussian model has the exact variance of a single feature but neglects the
correlations between qubits, and its gradients are those of the noisy loss,
without the additional shot noise of parameter-shift estimates. We train with $N_s=10^3$ at $L=1$ to 4, and with
$N_s=10^2$ and $10^4$ at $L=4$. We define the operational shot cost of a
trained model as the smallest $N_s$ at which its mean test accuracy lies within
0.01 of its accuracy with exact expectation values, interpolated in $\log N_s$
on a grid from 10 to $10^8$, and we report its geometric mean over the
initializations. It is undefined for three exactly trained models at $L=1$,
which we omit: the diverged Pauli run, and one noiseless and one depolarizing
model whose accuracy stays more than 0.01 below its exact value up to $10^8$
shots.

\subsection{Three-qubit variational eigensolver}
\label{sec:model_vqe}

Our third model is the eigensolver of Ref.~\cite{vanrossum2026biased} for the
periodic transverse-field Ising chain on three qubits,
\begin{equation}
H=-J\sum_{i=1}^{3}Z_iZ_{i+1}-h\sum_{i=1}^{3}X_i ,
\label{eq:tfim}
\end{equation}
with $J=1$, $h=0.5$, and $Z_4\equiv Z_1$, whose ground-state energy is
$E_0=-3.2321$. A trainable $R_Y$ on each qubit is followed by four blocks.
Each block applies $R_{ZZ}(\phi)$ to the three pairs and a trainable $R_X$ to
each qubit, and each two-qubit gate is decomposed as
\begin{equation}
R_{ZZ}(\phi)=\mathrm{CNOT}_{i,i+1}\,\bigl[I\otimes R_Z(\phi)\bigr]\,
\mathrm{CNOT}_{i,i+1} .
\label{eq:rzz}
\end{equation}
We compare two noise placements with the same number of insertions. In the
\textit{inside} placement of Ref.~\cite{vanrossum2026biased}, the channel acts
on both qubits after each CNOT. In the \textit{after} placement, it acts twice
on both qubits after the complete $R_{ZZ}$. Single-qubit gates are noiseless.
We minimize the energy and report the relative error,
\begin{equation}
E(\theta)=\Tr\bigl[H\rho(\theta)\bigr],\qquad
\epsilon=\frac{E(\theta)-E_0}{\lvert E_0\rvert},
\label{eq:vqe_energy}
\end{equation}
with Adam (learning rate 0.05, 500 steps) from 50 initializations per point,
drawn uniformly from $[0,2\pi)$, with every channel of a point starting from
the same parameters. Here and below we denote the noise strength of the
eigensolver by $\gamma$, the $p$ of Table~\ref{tab:channels}. For the inside
placement, our implementation reproduces the energies of the
Qiskit~\cite{javadiabhari2024quantum} code that the authors of
Ref.~\cite{vanrossum2026biased} deposited openly~\cite{vanrossum2026code} to $4\times10^{-14}$ for the
noiseless circuit, AD, reversed AD, and both twirls, at
$\gamma\in\{0.01,0.05,0.1,0.2\}$.

\subsection{Statistics}

Classifier results are given as the mean and standard deviation (SD) over
eight initializations with fixed data. All runs share one test set, so a
comparison between two conditions is also uncertain because the test set is
finite. For the comparisons on which our conclusions rest, we therefore quote a
95\% confidence interval (CI) from a two-level bootstrap~\cite{efron1994introduction} that resamples both
the initializations, jointly for the two conditions since they share initial
parameters, and the test images. The hypothesis tests behind every comparison
are collected in Appendix~\ref{app:stats}.

\section{Output scale}
\label{sec:scale}

A channel acts on a variational model first through the scale of its output.
For depolarizing noise on one qubit this is its only effect. Between AD and its
Pauli twirl, which share the contraction matrix $T$, the difference in scale is
itself an effect of the non-unital vector $\mathbf t$ (Sec.~\ref{sec:cost}).
Here we show what a trainable output scale removes and what it costs.

\subsection{Scale and sampling cost}

Suppose first that each noisy feature is an affine function of the noiseless one,
\begin{equation}
\tilde z_j=s\,z_j+b_j .
\label{eq:affine}
\end{equation}
Then a trainable affine readout realizes on
$\tilde{\mathbf z}$ every classifier it realizes on $\mathbf z$, and the noise
has no effect on the model class. For a single qubit under depolarizing
noise this holds exactly, since the channel commutes with every unitary: with
$N$ insertions,
\begin{equation}
\tilde f=\lambda_d^{N}f .
\label{eq:depol_scale}
\end{equation}
For anisotropic or non-unital
channels, and on several qubits, the noisy model class is not a rescaled copy
of the noiseless one, and whether the scale accounts for the differences is an
empirical question.

The rescaling is not free on hardware, as we now show. A feature estimated
from $N_s$ shots, $\hat z_j$, has the variance
\begin{equation}
\mathrm{Var}(\hat z_j)=\frac{1-z_j^2}{N_s}\le\frac1{N_s} .
\label{eq:shotvar}
\end{equation}
The number of shots is a leading cost of variational
algorithms~\cite{wecker2015progress,gonthier2022measurements}. The readout
divides the feature by its scale, so it amplifies the shot noise by the same
factor. Keeping the
class scores at a fixed precision therefore requires a number of shots proportional
to the inverse square of the feature scale. We report the shot factor
\begin{equation}
\kappa(\mathcal N)=\biggl(\frac{\sigma(\text{None})}{\sigma(\mathcal N)}\biggr)^{2},
\label{eq:kappa}
\end{equation}
where $\sigma$ is the mean over features and over the initializations of the standard deviation of $z_j$
over the training images at the trained parameters. For the single-qubit fits
we use $1/\lvert s\rvert$, with $s$ the median fitted output scale, in place of
$\sigma$, so that $\kappa=(s/s_{\rm None})^2$. In the language of
error mitigation, the trained readout performs a
rescaling~\cite{vovrosh2021simple,urbanek2021mitigating}, which cannot
remove an exponential concentration without an exponential number of
shots~\cite{wang2024errormitigation}. In Sec.~\ref{sec:shots} we compare
$\kappa$ with the operational shot cost defined in Sec.~\ref{sec:model_clf}.

\subsection{Single-qubit fits}
\label{sec:vr}

\begin{figure*}[t]
\centering
\includegraphics[width=\textwidth]{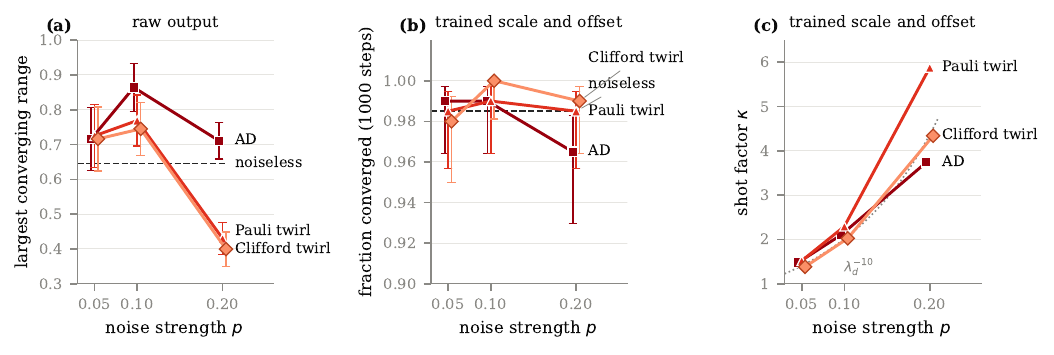}
\caption{Single-qubit re-uploading fits of Ref.~\cite{vanrossum2026biased}
(200 targets). (a) Largest converging output range of the raw model
$f=\langle Z\rangle$, with a free target offset as in
Ref.~\cite{vanrossum2026biased}. (b) Fraction of targets of range $R=1$ that
converge within 1000 steps when an output scale $s$ and offset $b$ are
trained. (c) Shot factor $\kappa=(s/s_{\rm None})^2$ from the median fitted
scale; the dotted line is the exact depolarizing value $\lambda_d^{-10}$. Bars
are 95\% confidence intervals of the mean over targets in (a) and Wilson 95\%
intervals in (b).}
\label{fig:vr}
\end{figure*}

With the raw output and a free target offset, we recover the ordering of
Ref.~\cite{vanrossum2026biased} at $p\ge0.1$ [Fig.~\ref{fig:vr}(a)]: at
$p=0.1$ the largest converging range is 0.864 for AD, against 0.769 and 0.745
for the Pauli and Clifford twirls. The metric is not monotone in the noise,
however. The noiseless model reaches only 0.646, below every noisy channel at
$p\le0.1$, so the metric measures convergence within the 150-step budget as
much as the range the model can reach; with targets centered on zero, the
ordering at $p=0.1$ reverses (Appendix~\ref{app:stats}).

With a trained scale and offset, we find that the difference disappears
[Fig.~\ref{fig:vr}(b)]. At $R=1$, AD and the Pauli twirl converge on 96.5 to
99.0\% and on 98.5 to 99.0\% of the targets within 1000 steps at $p=0.05$ to
0.2, the noiseless model on 98.5\%, and none of the paired differences is
significant; within 150 steps the only significant difference favors the Pauli
twirl (Appendix~\ref{app:stats}). What the twirl changes is the scale
[Fig.~\ref{fig:vr}(c)]. At $p=0.2$ the Pauli twirl needs 1.24 to 1.57 times
the shots of AD at the two ranges, and the Clifford twirl, whose shot factor
follows the exact depolarizing value $\lambda_d^{-10}$, 0.92 to 1.16 times. On
one qubit the cost of twirling is therefore modest; it becomes large in the
classifier, where it compounds over qubits and layers.
The same scale accounts for the smaller gradient magnitudes reported in
Ref.~\cite{vanrossum2026biased}, which notes that under Pauli noise the loss
gradients carry the attenuation factors of the output. For a rescaled output,
$\partial(sf)/\partial\theta=s\,\partial f/\partial\theta$, so a trained $s$
restores the gradient magnitude, but not the signal-to-noise ratio of
shot-estimated gradients, whose noise it amplifies by the same factor;
estimating the gradients to fixed precision costs the shot factor $\kappa$.

\subsection{Classifier}
\label{sec:clf}

\begin{table*}[t]
\caption{Test accuracy of the classifier on 3037 MNIST images (mean $\pm$ SD
over eight initializations) against the number of re-uploading layers $L$.
The raw readout is trained for 100 steps and, at $L=4$, for 1000 steps; the
standardized readout of Eq.~\eqref{eq:readout} for 100 steps.
$^\dagger$One of the eight runs diverged (test accuracy 0.382); the other seven
lie between 0.860 and 0.865.}
\label{tab:clf}
\begin{ruledtabular}
\begin{tabular}{lccccccccc}
 & \multicolumn{5}{c}{raw readout} & \multicolumn{4}{c}{standardized readout}\\
 & $L=1$ & $L=2$ & $L=3$ & $L=4$ & $L=4$, 1000 steps & $L=1$ & $L=2$ & $L=3$ & $L=4$\\
\colrule
None & $.861\pm.003$ & $.866\pm.005$ & $.865\pm.009$ & $.863\pm.004$ & $.862\pm.007$ & $.860\pm.005$ & $.869\pm.007$ & $.868\pm.006$ & $.869\pm.005$\\
AD & $.842\pm.010$ & $.835\pm.032$ & $.839\pm.005$ & $.841\pm.004$ & $.872\pm.003$ & $.861\pm.003$ & $.872\pm.003$ & $.877\pm.004$ & $.876\pm.005$\\
Pauli & $.847\pm.003$ & $.817\pm.039$ & $.787\pm.044$ & $.715\pm.057$ & $.837\pm.013$ & $.802\pm.170^\dagger$ & $.873\pm.005$ & $.869\pm.009$ & $.870\pm.007$\\
Depol & $.851\pm.003$ & $.814\pm.041$ & $.778\pm.034$ & $.695\pm.041$ & $.826\pm.012$ & $.860\pm.005$ & $.869\pm.006$ & $.869\pm.010$ & $.870\pm.004$\\
\end{tabular}
\end{ruledtabular}
\end{table*}

\begin{figure*}[t]
\centering
\includegraphics[width=\textwidth]{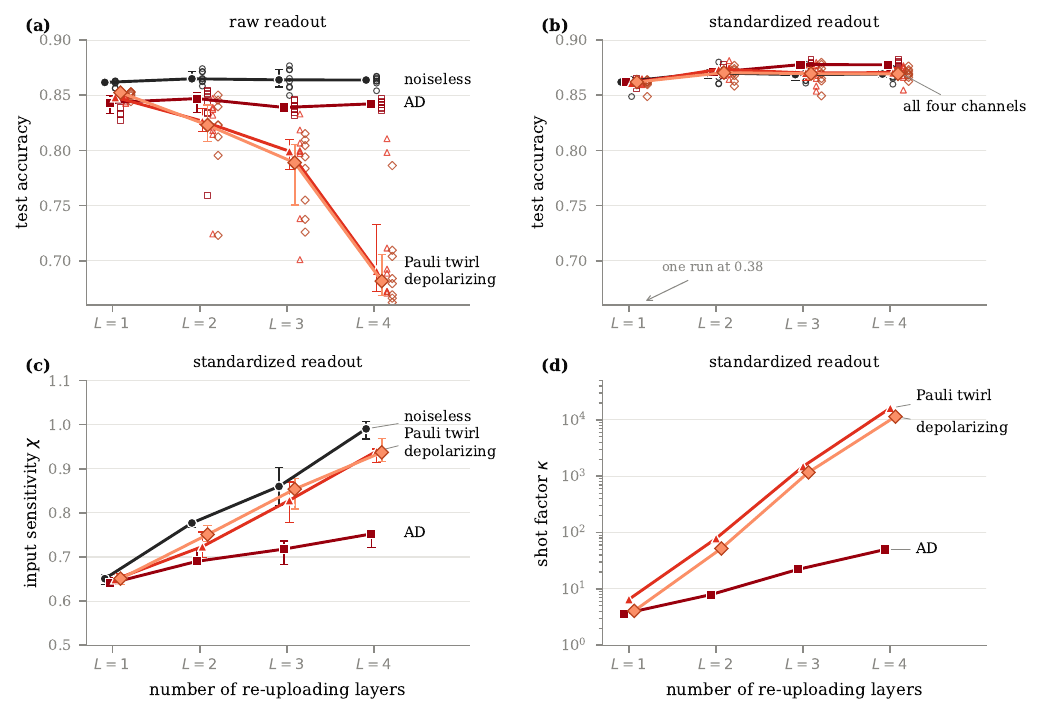}
\caption{Four-qubit classifier on MNIST against the number of re-uploading
layers. (a) Test accuracy with the raw readout and (b) with the
standardized readout, both after 100 steps. (c) Input sensitivity and (d)
shot factor $\kappa$ of Eq.~\eqref{eq:kappa} with the standardized readout.
Panels (a)--(c) show the median and interquartile range over eight
initializations, with each run as a hollow marker; one Pauli run at $L=1$ in
(b) diverged and lies below the axis.}
\label{fig:clf}
\end{figure*}

With the raw readout and 100 training steps, we find that the unital channels
appear to lose trainability with depth [Table~\ref{tab:clf}, Fig.~\ref{fig:clf}(a)].
Between $L=1$ and $L=4$ the test accuracy falls from 0.847 to 0.715 for the
Pauli twirl and from 0.851 to 0.695 for depolarizing noise, and the training
accuracy falls with it (to 0.700 and 0.672), while AD stays at 0.84 and the
noiseless circuit at 0.86. A reading in terms of trainability would conclude
that the non-unital term keeps the deep circuit trainable.

We find that the readout explains it. Trained for 1000 steps with the same raw readout,
the unital channels recover to 0.837 and 0.826 at $L=4$, while the norm of
their readout weights grows from about 19 at step 100 to 106 and 109 at step
1000; for the noiseless circuit it grows from 8 to 20. With the standardized
readout, every channel reaches the accuracy of the noiseless circuit at every
depth [Fig.~\ref{fig:clf}(b)]. The Pauli and depolarizing channels are then
indistinguishable from it: at $L=4$ the differences are $+0.001$ (95\% CIs
$[-0.007,+0.009]$ and $[-0.006,+0.008]$), and no
difference at $L=2$ or 3 exceeds 0.004. At 100 steps the raw readout cannot
grow its weights fast enough to use features whose scale the unital channels
have shrunk to 4 and 5 percent of the noiseless scale after 100 steps (and to
below 1 percent before training); the circuit itself remains trainable.
At $L=1$ the depolarizing channel provides an exact check. Each feature then
crosses all three noise insertions as a weight-one Pauli string, so the
channel multiplies every feature by the same factor for every $\theta$,
\begin{equation}
z_j^{\rm Depol}(\mathbf a,\theta)=\lambda_d^{3}\,z_j^{\rm None}(\mathbf a,\theta) .
\label{eq:depol_L1}
\end{equation}
The standardized readout removes a uniform scale, and the depolarizing and
noiseless classifiers indeed make identical predictions on every test image.
AD is not a rescaling, and it ends slightly above the noiseless circuit at
$L=3$ and 4 ($+0.009$, CI $[+0.003,+0.015]$, and $+0.008$, CI
$[+0.002,+0.013]$), but not at $L\le2$ (Sec.~\ref{sec:cost}).
The same holds on Fashion-MNIST at $L=4$. With the raw readout, the unital
channels are 16 and 17 points below the noiseless circuit (0.765) and AD is 5
points below it. With the standardized readout, the unital channels are not
resolved from the noiseless circuit, and AD is above it by 0.017, CI
$[+0.007,+0.029]$ (Appendix~\ref{app:stats}). The shot factors are 34 for AD
and $9.9\times10^3$ and $8.2\times10^3$ for the Pauli and depolarizing
channels.

What survives, as we show in Fig.~\ref{fig:clf}(d), is the feature scale. At the end of
training with the standardized readout, the shot factor grows by a factor of
about 14 per layer for the unital channels, from 4.1 at $L=1$ to
$1.1\times10^4$ at $L=4$ for depolarizing noise, and to $1.6\times10^4$ for the
Pauli twirl. For AD it grows by a factor of about 2.4 per layer, from 3.6 to
50. The shot factor depends on the training objective and is not a property
of the channel alone. With the raw readout the optimizer moves the parameters
toward less contracted features: after 1000 steps at $L=4$ the factors are 11
for AD, 320 for the Pauli twirl, and 470 for depolarizing noise. The measured
shot costs follow $\kappa$ for the unital channels but not for AD
(Sec.~\ref{sec:shots}).

\section{Gauge freedom of the damping direction}
\label{sec:gauge}

\subsection{When the damping direction is a gauge}
\label{sec:gauge_def}

Amplitude damping toward $\ket{1}$ instead of $\ket{0}$, $\AD^{\rm flip}$ of
Eq.~\eqref{eq:adflip}, has the same contraction matrix $T=\mathrm{diag}(c,c,c^2)$ as $\AD$
and the opposite non-unital vector, $\mathbf t=(0,0,-p)$. Whether a circuit can tell the two apart is not a property of the
channel but of where the circuit lets a Pauli frame travel.

Write the reversed circuit as $\Phi^{\rm rev}=E^{\rm rev}_m\circ\dots\circ E^{\rm rev}_1$,
where the elements $E^{\rm rev}_k$ are its gates, encodings, and noise layers. We insert
$g_k\circ g_k=\mathrm{id}$ at every cut $k$ between the elements. Each $g_k$ is a Pauli frame~\cite{knill2005quantum}: a Pauli
conjugation, or its composition with complex conjugation in the computational basis,
\begin{equation}
g_k(\rho)=
\begin{cases}
P_k\,\rho\,P_k & \text{(unitary type)},\\
P_k\,\bar\rho\,P_k & \text{(antiunitary type)},
\end{cases}
\label{eq:frame}
\end{equation}
with all $g_k$ of one type. The reversed circuit becomes
\begin{align}
\Phi^{\rm rev}&=g_m\circ\tilde E_m\circ\dots\circ\tilde E_1\circ g_0,
\label{eq:framed}\\
\tilde E_k&=g_k\circ E^{\rm rev}_k\circ g_{k-1} .
\label{eq:framed_k}
\end{align}

\begin{proposition}[Frame gauge]\label{prop:gauge}
Reversing the damping is a reparametrization of the model,
\begin{equation}
y^{\mathrm{rev}}(x;\theta)=\eta\,y(x;\theta')
\label{eq:reparam}
\end{equation}
for all inputs $x$, with $\eta=\pm1$, whenever a frame assignment of one type exists such that
(i) at every noise location the frame acts as $X$ or $Y$ on each noisy qubit;
(ii) every fixed gate maps the incoming frame to a Pauli frame;
(iii) every data-encoding gate is left invariant;
(iv) every trainable element is mapped back into its own family;
(v) the input state is absorbed by the first element on each qubit, and the final frame
maps each measured observable to $\eta$ times itself, with $\eta=-1$ allowed only if the readout
has a learnable sign.
\end{proposition}
\noindent The proof is given in Appendix~\ref{app:gauge}.

Propagating Pauli operators through a parametrized circuit is the $\sigma$-pulse
method of Fontana \textit{et al.}~\cite{fontana2022nontrivial}, who used it to
find parameter symmetries of a single model. They showed that these symmetries
survive unital Pauli noise and can be broken by non-unital noise such as
amplitude damping. Where a propagated frame crosses the channel as $X$ or $Y$,
it maps $\AD$ onto $\AD^{\rm flip}$, and this is how the symmetry breaks.
Proposition~\ref{prop:gauge} reads the same propagation as a map between two
noise models, adds the antiunitary frames, and states the conditions on the
encoding, the input, and the readout under which the map is a
reparametrization. Noise can also lift degeneracies among the parameters of
overparametrized circuits~\cite{fontana2021evaluating}.

The antiunitary case contains spin time reversal,
\begin{equation}
\Theta=\bigotimes_q(-iY_q)\,K ,
\label{eq:theta}
\end{equation}
with $K$ complex conjugation in the computational basis. It reverses every Pauli
matrix and the imaginary unit,
\begin{equation}
\Theta\,\boldsymbol\sigma_q\,\Theta^{-1}=-\boldsymbol\sigma_q,\qquad
\Theta\,i\,\Theta^{-1}=-i ,
\label{eq:theta_sigma}
\end{equation}
so every single-qubit rotation is time-reversal invariant, while the damping is
reversed because the Kraus operators of $\AD$ are real,
\begin{equation}
\Theta R_{\mathbf n}(a)\,\Theta^{-1}=R_{\mathbf n}(a),\qquad
\Theta\,\AD\,\Theta^{-1}=\AD^{\rm flip},
\label{eq:theta_inv}
\end{equation}
for any axis $\mathbf n$. Condition (iii) therefore never fails for single-qubit encodings, whatever their
axis: only multi-qubit fixed gates, the placement of noise relative to them, and the two
boundaries can break the gauge. In our classifier (noise after the whole entangling ring,
a general trainable layer before it, an affine readout) all five conditions hold for
$R_X$, $R_Y$, $R_Z$ and a generic encoding axis alike, and the reversal is exact
reparametrization to within $3.3\times10^{-16}$ (Table~\ref{tab:gauge}, at $n=3$
and $L=2$; with the explicit maps of App.~\ref{app:gauge}, to
$9.4\times10^{-16}$ for $n=2$ to 4 and $L=1$ to 4). A unitary Pauli frame suffices only when it
commutes with the encoding generator and still flips the damping, which singles out $X$ for
$R_X$ and $Y$ for $R_Y$; for $R_Z$ or a generic axis only the antiunitary frame works. An
encoding choice therefore never protects the damping direction from being a gauge. The damping-direction comparison of
Sec.~\ref{sec:gauge_clf} therefore carries no information about the sign of the
non-unital term.

\begin{corollary}[Deterministic frames are gauge]\label{cor:det}
Under the conditions of Proposition~\ref{prop:gauge}, a deterministic relabelling of the
noise by a Pauli frame or by time reversal leaves the reachable model class unchanged. If, in
addition, $\theta\mapsto\theta'$ preserves the distribution of the initial parameters and the
optimizer is equivariant under it, the trained models under $\AD$ and $\AD^{\rm flip}$ are
identically distributed.
\end{corollary}

\noindent Both conditions hold for our classifier. For each frame of
Table~\ref{tab:gauge}, every Euler angle of the trainable layers maps as
$\theta_k\mapsto\pm\theta_k+m_k\pi$ with $m_k\in\{0,1\}$, and the
readout as $W\mapsto-W$, $b\mapsto b$. These maps leave the initialization invariant,
since the model depends on each angle only modulo $2\pi$. The standardized readout maps
$\boldsymbol\mu\mapsto-\boldsymbol\mu$ and $\boldsymbol\sigma\mapsto\boldsymbol\sigma$,
and Adam without weight decay is equivariant under sign flips and translations of the
coordinates. Trained classifiers under $\AD$ and $\AD^{\rm flip}$ are therefore identically
distributed over initializations; the same holds with shot noise, which is symmetric under
$z\mapsto-z$. In the eigensolver of Ref.~\cite{vanrossum2026biased} with the channel
moved after each complete $R_{ZZ}$ (Sec.~\ref{sec:gauge_vqe}), the frame changes only the
initial $R_Y$ angles, $y\mapsto\pi-y$, which preserves their initialization $U(0,\pi)$. The
optimizer of Ref.~\cite{vanrossum2026biased}, COBYLA, builds its first simplex along the
positive coordinate axes and is not equivariant under this reflection, so there only the
reachable energies are guaranteed to coincide. With the channel after each CNOT, as in
Ref.~\cite{vanrossum2026biased}, no frame exists (Table~\ref{tab:gauge}).

\begin{corollary}[The Pauli twirl is not frame-equivalent to AD]\label{cor:rand}
Pauli twirling replaces $\AD$ by a convex mixture of frame-equivalent channels,
\begin{equation}
\AD_{\rm Pauli}=\frac14\sum_{P\in\{I,X,Y,Z\}}P\,\AD\,P ,
\label{eq:twirl}
\end{equation} which is Eq.~\eqref{eq:twirl_mix}. It is not itself frame-equivalent to $\AD$: every (anti)unitary
Pauli frame preserves $\lVert\mathbf t\rVert$, which is $p$ for $\AD$ and $0$ for the twirl.
\end{corollary}

\begin{table*}[t]
\caption{Frame gauge for reversed amplitude damping, from an exhaustive search over all
Pauli frames of each type ($4^n$ strings; $n\le3$), with $p=0.3$. ``Gauge'' means a frame
assignment satisfying (i)--(v) exists; the explicit identity
$y^{\mathrm{rev}}(x;\theta)=\eta\,y(x;\theta')$ is then checked on random parameters and inputs, and
the largest deviation is quoted. ``None'' means no frame of that type satisfies (i)--(v). The eigensolver rows place the channel once after each whole $R_{ZZ}$ or after each CNOT; applying it twice after each $R_{ZZ}$, as in Fig.~\ref{fig:vqe}(a), leaves the frame assignment unchanged.}
\label{tab:gauge}
\begin{ruledtabular}
\begin{tabular}{lll}
model & Pauli frame & time reversal / antiunitary\\
\colrule
classifier, $R_X$ encoding, noise after ring  & gauge, $2.2\times10^{-16}$ & gauge, $3.3\times10^{-16}$\\
classifier, $R_Y$ encoding                     & gauge ($Y$ frame), $3.3\times10^{-16}$ & gauge, $2.2\times10^{-16}$\\
classifier, $R_Z$ encoding                     & none       & gauge, $2.2\times10^{-16}$\\
classifier, generic axis                       & none       & gauge, $3.3\times10^{-16}$\\
classifier, noise inside the ring              & none       & none\\
classifier, encoding before first trainable layer & none    & none\\
single-qubit regression, $f=\langle Z\rangle$  & none       & none\\
\quad with a learnable sign                    & gauge, $\eta=-1$, $1.1\times10^{-16}$ & gauge, $\eta=-1$, $2.8\times10^{-16}$\\
eigensolver, noise after each $R_{ZZ}$         & gauge, $2.8\times10^{-16}$ & gauge, $4.4\times10^{-16}$\\
eigensolver, noise after each CNOT            & none       & none\\
\end{tabular}
\end{ruledtabular}
\end{table*}

The single-qubit fits of Ref.~\cite{vanrossum2026biased} (Sec.~\ref{sec:vr};
general single-qubit blocks, $R_X$ encoding, noise after every block, output
$f=\langle Z\rangle$ used directly) fail only condition (v): the frame left
after the last channel flips the sign of $\langle Z\rangle$ and nothing
absorbs it, so
\begin{equation}
f^{\mathrm{rev}}(x;\theta)=-f(x;\theta') .
\label{eq:vr_mirror}
\end{equation}
The two model classes are mirror images and differ as sets, since the last
channel confines the output under AD and under reversed AD to
\begin{equation}
f(x;\theta)\in[2p-1,\,1],\qquad f^{\mathrm{rev}}(x;\theta)\in[-1,\,1-2p] ;
\label{eq:vr_ranges}
\end{equation}
both upper bounds are attained. A target ensemble symmetric under $f\to-f$, or a
trainable output scale, makes the two channels statistically
indistinguishable.

\subsection{Classifier}
\label{sec:gauge_clf}

In the classifier of Fig.~\ref{fig:circuit}, time reversal maps each CNOT of
the ring to
\begin{equation}
\Theta\,\mathrm{CNOT}\,\Theta^{-1}=-\mathrm{CNOT}\,(Z\otimes X),
\label{eq:theta_cnot}
\end{equation}
which leaves a local Pauli residue on its input side. We can move this residue
before the ring and absorb it into the preceding trainable blocks; $\Theta\ket0$
is absorbed by the first block, and the sign of the features by the readout. Proposition~\ref{prop:gauge} therefore applies for
every encoding axis (Table~\ref{tab:gauge}). When the noise acts after each
CNOT of the ring, no frame of either type exists; the exhaustive search of
Table~\ref{tab:gauge} rules out both.

In the classifier with the raw readout, we find that AD$^{\rm flip}$ reaches $0.838\pm0.008$ against
$0.841\pm0.004$ for AD at $L=4$ (95\% CI of the difference
$[-0.008,+0.002]$), with 96.7\% of the
individual predictions in agreement. By Corollary~\ref{cor:det}, the two
conditions are identically distributed over initializations, so the expected
difference is exactly zero; predictions differ because paired runs start from
the same $\theta$ under both channels, not from $\theta$ and $\theta'$. The
comparison therefore has a known null answer. The two-level interval includes
zero (paired $t$ test, $P=0.21$), whereas an interval that resamples only the
test images, $[-0.0056,-0.0009]$, excludes it: resampling the test set alone
understates the uncertainty.

When the noise acts after each CNOT of the ring, no frame exists, and the two
directions define different model classes. Trained with the standardized
readout at $L=4$, AD$^{\rm flip}$ reaches $0.867\pm0.009$ against
$0.873\pm0.004$ for AD, CI $[-0.013,+0.001]$; the two agree on 94.6\% of the
test images and in sensitivity (0.639 and 0.647). This variant applies the
channel more often, so we compare its two directions only with each other.
Breaking the gauge therefore allows a direction effect but does not guarantee
a large one; here the trainable layers and the readout may compensate for it,
whereas the eigensolver, with a fixed observable, cannot
(Sec.~\ref{sec:gauge_vqe}).

\subsection{Eigensolver}
\label{sec:gauge_vqe}

\begin{figure*}[t]
\centering
\includegraphics[width=\textwidth]{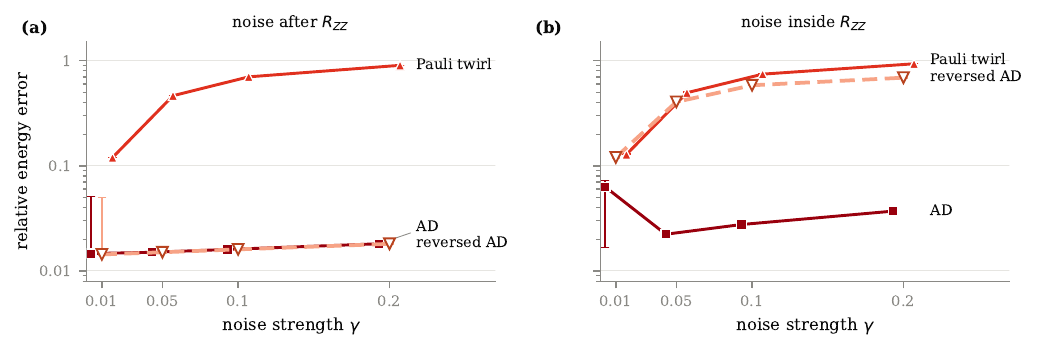}
\caption{Relative energy error of the three-qubit eigensolver against the
noise strength, for AD, reversed AD (hollow markers), and the Pauli twirl
(logarithmic vertical axis). Points are the median of 50 initializations and bars the
interquartile range. (a) The channel acts after the complete $R_{ZZ}$; AD and reversed
AD coincide. (b) The channel acts after each CNOT inside $R_{ZZ}$, as in
Ref.~\cite{vanrossum2026biased}.}
\label{fig:vqe}
\end{figure*}

In the eigensolver, the global flip $X^{\otimes3}$ leaves $H$ invariant. When
the noise follows the complete $R_{ZZ}$, $X^{\otimes2}$ commutes with the gate
and the flip reaches the initial $R_Y$ rotations, where it becomes a
reparametrization,
\begin{equation}
E_{\rm flip}(\theta)=E_{\AD}(\theta'),\qquad \theta'_q=\pi-\theta_q ,
\label{eq:vqe_gauge}
\end{equation}
where $\theta_q$ is the initial $R_Y$ angle on qubit $q$ and all other angles
are unchanged. We find that the two sides of Eq.~\eqref{eq:vqe_gauge} agree to
$1.4\times10^{-15}$ over 50 random parameter sets at every $\gamma$. Trajectories started from
mapped initial parameters end at energies that agree to $6\times10^{-15}$,
and the optimized AD and reversed-AD energies coincide [Fig.~\ref{fig:vqe}(a)].
The initialization of Ref.~\cite{vanrossum2026biased} is also invariant under
this map, but its optimizer, COBYLA~\cite{powell1994direct}, is not
(Sec.~\ref{sec:gauge_def}).

When the noise acts after each CNOT, the frame leaves a Pauli $X$ on the
target qubit between two noise channels, and the map fails: the energies of
the two channels at mapped parameters differ by 0.10 to 0.66 between
$\gamma=0.01$ and 0.2. The optimized energies separate
[Fig.~\ref{fig:vqe}(b)]. The median relative error of reversed AD exceeds
that of AD by 0.06, 0.38, 0.56, and 0.65 at $\gamma=0.01$, 0.05, 0.1, and 0.2
(Appendix~\ref{app:stats}); small initial angles, drawn from
$\mathcal N(0,0.1^2)$, give the same gaps within 0.01 at $\gamma\ge0.05$. The
best of the 50 initializations for
reversed AD lies within 0.01 of its median and above the median for AD at
every $\gamma$. L-BFGS-B~\cite{byrd1995limited} from six further random
initializations, run on an
independently written simulator, reaches the same minima at $\gamma=0.2$
(relative errors 0.0370 for AD and 0.6881 for reversed AD). The gap is
therefore a floor of the energy landscape, not a failure of the optimizer. This reproduces the direction effect of
Ref.~\cite{vanrossum2026biased} and locates its origin in the placement of
the noise inside the decomposed gate. On hardware whose native two-qubit gate
is the CNOT, the damping acts at this location.

On a bipartite coupling graph the effect cannot depend on the sign of the
coupling. Flipping the spins of one sublattice, $\prod_{i\in B}X_i$, maps
$J$ to $-J$ and commutes with the transverse field; it acts after the last
channel, where the final $R_X$ layer absorbs it ($x_i\mapsto x_i+\pi$ for
$i\in B$). On the open chain, with $B$ the middle qubit, the energies of the two
couplings agree at the mapped parameters to $2\times10^{-15}$ for every channel
and placement, and the shift preserves the uniform initialization. On the
three-site ring no such frame exists, since no product of Pauli operators flips
every bond of an odd cycle; we have not run the ring with $J<0$. With $J=\pm1$, $h=0.5$, and $\gamma=0.1$ on the open chain, reversed
AD reaches a median relative error of 0.367 against 0.063 for AD in both
cases, while the bond correlation $\langle Z_iZ_{i+1}\rangle$ of the optimized
AD state is $+0.89$ and $-0.89$. At $h=2$ on the ring, where the ground state
is paramagnetic, reversed AD still trails AD with the damping inside the gate
(0.59 against 0.29 at $\gamma=0.1$), and the two coincide with the damping
after it. Which direction the trapped frame favors is thus set by the circuit,
not by the ground state. Nor does the advantage of AD over its twirl rest on
the proximity of the damping's fixed point $\ket{000}$ to the ground state:
at $h=2$ on the ring, $\ket{000}$ and every state the final $R_X$ layer makes of
it have a relative error of at least 0.54, yet AD reaches 0.29 and 0.23 with
the damping inside and after the gate, against 0.66 and 0.59 for the twirl.
(The antiferromagnetic open chain does not test this: the sublattice flip maps
it onto the ferromagnetic chain, where $\ket{000}$ lies at 0.17.) On the ring, in both placements, the Pauli
twirl gives the largest error, 0.70 and 0.74 at $\gamma=0.1$. Unlike a classifier
output, the energy cannot be rescaled freely, so here the twirl costs accuracy
unless a mitigation step restores the scale~\cite{vovrosh2021simple}.

\section{What the non-unital term changes}
\label{sec:cost}

The output scale of Sec.~\ref{sec:scale} is not independent of the
non-unital term. The Pauli twirl has exactly the contraction matrix of AD and
differs only by $\mathbf t=\mathbf 0$, so the larger feature scale under AD is
itself an effect of $\mathbf t$. Here we show how $\mathbf t$ sets that scale and
the sampling cost that follows from it (Sec.~\ref{sec:shots}), and what it
changes beyond them.

\subsection{Sampling cost}
\label{sec:shots}

\begin{figure*}[t]
\centering
\includegraphics[width=\textwidth]{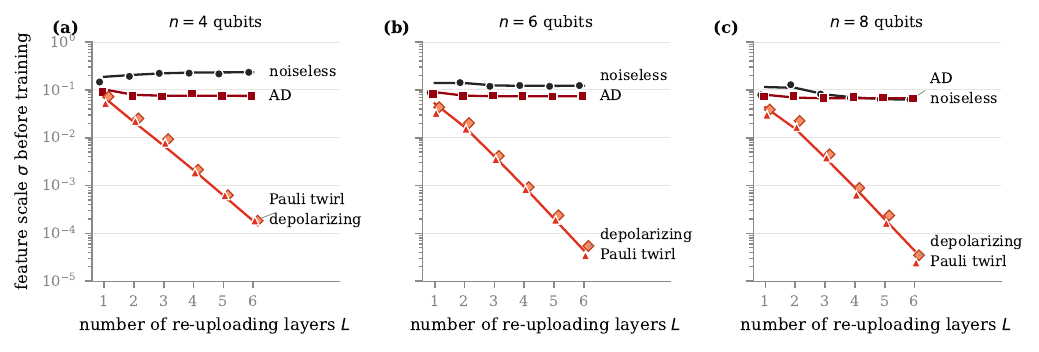}
\caption{Feature scale $\sigma$ before training against the number of
re-uploading layers for (a) four, (b) six, and (c) eight qubits, with $n$
principal components as inputs. Markers are the root-mean-square scale over
five parameter draws, the same at every depth, and 300 training images; lines are
the exact average over parameters for the noiseless circuit, AD, and the Pauli
twirl (Appendix~\ref{app:width}).}
\label{fig:width}
\end{figure*}

\begin{figure*}[t]
\centering
\includegraphics[width=\textwidth]{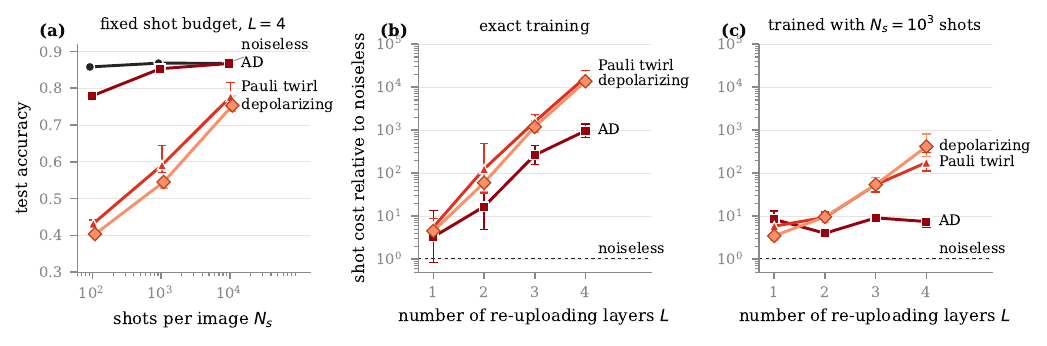}
\caption{Finite shots in the MNIST classifier. (a) Test accuracy at $L=4$ of
models trained and tested with $N_s$ shots per image (median and
interquartile range over eight initializations). (b), (c) Operational shot
cost relative to the noiseless circuit (geometric mean over initializations,
with 95\% bootstrap intervals over initializations) for models trained with
exact expectation values and with $N_s=10^3$ shots.}
\label{fig:shots}
\end{figure*}

Before training, the feature scale separates the channels with depth and
width (Fig.~\ref{fig:width}). The unital channels shrink it by a nearly
constant factor per layer at every width. Under AD it settles after two or
three layers on a floor of 0.067 to 0.075 for $n=4$ to 8. In the exact average
over parameters (Appendix~\ref{app:width}), the ratio of the AD and
Pauli-twirl scales grows by a factor of 3.3 per layer at $n=4$ and 4.5 for
$n\ge6$, so its square, which sets the relative shot cost, grows by 11 and 20
per layer. At $n=4$ the squared ratio is 2.1 at $L=1$ and $1.3\times10^3$ at
$L=4$. At $p=0.3$ the floor has a closed form. The damping after the last entangler
branches the $Z$ component of each feature into the identity, which reaches the
readout whatever the earlier layers do, and contributes
\begin{equation}
\delta z_j=p(1-p)\bigl[(1-p)\,r_z\cos a_j-c\,r_y\sin a_j\bigr]
\label{eq:floor}
\end{equation}
to the feature, where $(r_x,r_y,r_z)$ is the direction into which the final
trainable block rotates $Z_j$. The noiseless scale falls with width, as
expected for a scrambling circuit, but the floor changes little, so at $n=8$ and
$L\ge4$ the AD features are as large as the noiseless ones. This is
consistent with the effective shallowness of noisy
circuits~\cite{mele2026noise}; under AD the last layer alone sets the scale of
the features. It also agrees with the theory of noise-induced concentration:
unital noise concentrates local expectation values exponentially in
depth~\cite{wang2021noiseinduced,schumann2024emergence}, whereas under
non-unital noise the variance of local expectation values over random circuits
stays bounded below by the size of the non-unital term~\cite{mele2026noise}.
Equation~\eqref{eq:floor_var} is the counterpart of that bound for the variance
over inputs: branchings in different layers are uncorrelated after the average
over parameters, so $\sigma_{\AD}^2\ge\sigma_\infty^2$ at every depth. Both
bounds scale as $\lVert\mathbf t\rVert^2=p^2$ and do not decay with depth; at
weak damping the branchings of earlier layers raise the floor above this
bound. The same effective shallowness allows local expectation
values of such circuits to be estimated classically on
average~\cite{mele2026noise}, and for Pauli noise also by truncating Pauli
paths~\cite{fontana2023classical}. For quantum kernels, noise likewise
concentrates the values over different inputs, so that polynomially many
shots give a model independent of the input~\cite{thanasilp2024exponential}.
In reservoir computing, the state of a contractive reservoir is the maximally
mixed state for every input sequence if and only if the reservoir map is
unital~\cite{martinezpena2023quantum}.

At finite temperature the damping of a qubit with splitting $\hbar\omega$
relaxes toward the thermal population $N=(e^{\hbar\omega/k_BT}+1)^{-1}$, and at
fixed $p$, that is, at fixed measured $T_1$, $t_z=p(1-2N)=p\tanh(\hbar\omega/2k_BT)$.
At fixed coupling to the bath, $1/T_1$ itself grows as $\coth(\hbar\omega/2k_BT)$,
so at weak damping $t_z$ is independent of temperature to first order in $p$,
and temperature acts through the contraction. The last-layer term,
Eq.~\eqref{eq:floor}, is linear in $t_z$, so it scales exactly as $1-2N$.
Strings of higher Pauli weight can branch into the identity on several qubits
and add higher powers of $t_z$, but they are small: in the exact average over
parameters at $n=4$, the branching part of the feature scale follows
$\lvert1-2N\rvert$ within 0.4\% for every $N$ and $L\le8$ at $p=0.3$ (within
0.2\% at $p=0.1$), and it is exactly symmetric under $N\leftrightarrow1-N$, as
the frame requires. The floor thus vanishes linearly in $1-2N$ as the temperature rises, and
the Pauli twirl is the infinite-temperature end of this family.

After training with exact expectation values, the measured shot costs follow
$\kappa$ for the unital channels but not for AD [Fig.~\ref{fig:shots}(b)]. At
$L=4$ on MNIST the operational cost relative to the noiseless circuit is
$1.7\times10^4$ for the Pauli twirl and $1.4\times10^4$ for depolarizing noise,
against $\kappa=1.6\times10^4$ and $1.1\times10^4$, and 967 for AD, against
$\kappa=50$. The Pauli twirl then needs 17 times the shots of AD on MNIST and
36 times on Fashion-MNIST. The excess of AD comes from the readout. The shot
factor averages the spreads of the features, whereas the readout amplifies the
shot noise of each feature by its own weight. The readout-weighted noise gain
$G=\bigl\langle\sum_j\lVert W_{:,j}\rVert^2(1-z_j^2)/\sigma_j^2\bigr\rangle$,
averaged over the test images, reproduces the measured ratios to within a
factor of 1.6 for $L\ge2$. Exact training lets the AD readout rely on features whose
spread is small: at $L=4$ the geometric mean of the smallest feature spread is
0.0065 for AD, against 0.24 for the noiseless circuit.

Training under shot noise changes this picture [Fig.~\ref{fig:shots}(a) and
(c)]. With $N_s$ shots in training and test, AD stays on average within 1.5 percentage points of the
noiseless accuracy for $N_s\ge10^3$ at every depth, whereas at $N_s=10^3$ and
$L=4$ the Pauli and depolarizing channels lose 27 and 33 points, and at
$N_s=10^4$ still 9 and 12 points. At $N_s=10^2$ AD also loses 8 points. Part
of the loss arises in training. Evaluated with exact expectation values, the
twirled models trained with $10^3$ shots reach 0.762 and 0.758, against 0.870
when trained exactly, while AD reaches 0.864, against 0.876. The operational
cost relative to the noiseless circuit then stays between 4 and 10 for AD at
every depth, close to its shot factor (7.5 against 8.5 at $L=4$). For the
twirls it still grows by a factor of 3 to 5 per layer, to 177 and 417 at
$L=4$. Training under shot noise moves the circuit toward features whose
spread exceeds the shot noise. At $L=4$ and $N_s=10^3$ the geometric mean of
the smallest feature spread is 0.075 for AD, against 0.0065 under exact
training, above $1/\sqrt{N_s}\approx0.03$ and at the floor of
Fig.~\ref{fig:width}; for the twirls it stays at 0.005 and 0.003, below the
shot noise. At $L=1$ the cost of AD is
the largest of the three channels (8.4, against 5.9 and 3.5), with all costs
small; we do not have an explanation for this.

\begin{table}[t]
\caption{Dependence on the damping strength $p$ before training, from the
exact average over parameters: the depth $L^*$ at which
$(\sigma_{\AD}/\sigma_{\rm Pauli})^2$ first reaches 10, the floor
$\sigma_{\rm br,\infty}$ of the AD scale and its last-layer part $\sigma_\infty$
[Eq.~\eqref{eq:floor_var}], and the number of shots
$1/\sigma_{\rm br,\infty}^2$ that resolves the floor. The floor is estimated from 5000 ($n=4$) and 700 ($n=8$) random image pairs, with a sampling error of about 1\%.}
\label{tab:pdep}
\begin{ruledtabular}
\begin{tabular}{lccccc}
 & \multicolumn{4}{c}{$n=4$} & $n=8$\\
$p$ & $L^*$ & $\sigma_{\rm br,\infty}$ & $\sigma_\infty$ & $1/\sigma_{\rm br,\infty}^2$ & $L^*$\\
\colrule
0.001 & 1508 & 0.0017 & 0.0005 & $3.4\times10^5$ & --\\
0.003 & 453 & 0.0031 & 0.0014 & $1.0\times10^5$ & --\\
0.01 & 117 & 0.0067 & 0.0047 & $2.2\times10^4$ & 49\\
0.03 & 33 & 0.0155 & 0.0137 & $4.2\times10^3$ & 13\\
0.05 & 18 & 0.0235 & 0.0219 & $1.8\times10^3$ & 8\\
0.1 & 8 & 0.0408 & 0.0396 & 600 & 5\\
0.2 & 4 & 0.0647 & 0.0637 & 240 & 3\\
0.3 & 2 & 0.0757 & 0.0747 & 170 & 2\\
\end{tabular}
\end{ruledtabular}
\end{table}

\begin{figure*}[t]
\centering
\includegraphics[width=\textwidth]{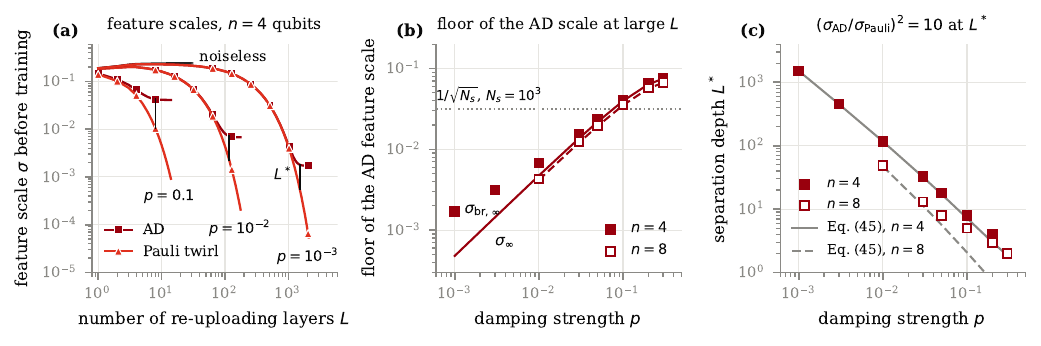}
\caption{Dependence of the separation on the damping strength, from the exact
average over parameters before training (5000 image pairs at $n=4$, 700 at
$n=8$). (a) Feature scale against depth at $n=4$ for AD (squares) and the
Pauli twirl (triangles) at three values of $p$, and for the noiseless circuit
up to $L=30$; vertical segments mark $L^*$. (b) Floor $\sigma_{\rm br,\infty}$
of the AD scale (markers) and its last-layer part $\sigma_\infty$
[Eq.~\eqref{eq:floor_var}, lines] for $n=4$ (filled, solid) and $n=8$ (hollow,
dashed); the dotted line is the shot noise $1/\sqrt{N_s}$ at $N_s=10^3$.
(c) Separation depth $L^*$ (markers) and Eq.~\eqref{eq:Lstar} (gray lines).
At $n=8$ the exact average was computed for $p\ge0.01$.}
\label{fig:sep}
\end{figure*}

The separation depends on the damping strength (Table~\ref{tab:pdep} and
Fig.~\ref{fig:sep}). At
weak damping the twirled scale decays per layer as $e^{-np}$, and the square of
the AD floor has two parts,
$\sigma_{\rm br,\infty}^2\approx\sigma_\infty^2+b_{\rm br}\,p$: the
last-layer term of Eq.~\eqref{eq:floor_var}, which is proportional to $p^2$,
and the branchings of the roughly $1/(np)$ earlier layers that the damping does
not yet suppress. The second part acts on the data dependence of the earlier
layers, which concentrates with width, so that for $p\le0.03$ the coefficient
$b_{\rm br}$ falls from $(1.8$--$2.7)\times10^{-3}$ at $n=4$ to
$(1.0$--$2.4)\times10^{-4}$ at $n=8$. The squared ratio of the AD and twirl scales, which is the ratio of
their shot costs, reaches 10 at the depth
\begin{equation}
L^*\approx\frac{1}{np}\,\ln\frac{3\,\sigma_{\rm None}}{\sigma_{\rm br,\infty}},
\label{eq:Lstar}
\end{equation}
where $\sigma_{\rm None}$ is the noiseless scale at large depth (0.24 at $n=4$
and 0.062 at $n=8$). At $n=4$ this agrees [Fig.~\ref{fig:sep}(c)] with the exact value within 1\% for
$p\le0.01$ and within one layer for larger $p$, and at $n=8$ within two layers
for $p\le0.03$; it grows roughly as $p^{-1}\ln(1/p)$. At fixed $p\le0.05$ it falls
faster than $1/n$ (from 117 at $n=4$ to 49 at $n=8$ for $p=0.01$), since the
noiseless scale in the logarithm concentrates with width faster than the
floor. Damping on superconducting hardware is of order $10^{-3}$ per
two-qubit gate, for gate times of 30 to 400~ns against $T_1$ of order
$100~\mu$s~\cite{krantz2019quantum}; in our model one of the two insertions per
layer follows a ring of $n$ CNOTs, so $p$ per insertion corresponds to several
gate durations. At
$p=10^{-3}$ the separation needs about 1500 layers at $n=4$, and resolving the
AD floor needs $3\times10^5$ shots. The difference in feature scale between AD
and its twirls is thus a property of strong damping or of very deep circuits.
Trained classifiers follow this picture (Table~\ref{tab:pdep_trained}). At
$L=4$ and $p=0.05$, well below $L^*=18$, AD and its twirls reach the same
accuracy at nearly the same shot cost, in agreement with $\kappa$. At $p=0.1$,
closer to $L^*=8$, the Pauli twirl needs 1.6 times the shots of AD, between
the ratio 1.4 of the trained feature scales and the squared ratio 1.7 before
training, and AD is more accurate than the Pauli twirl by 0.009, CI
$[+0.002,+0.016]$. At $p=0.1$ and $L=L^*=8$ the channels separate: the twirl
needs 4.1 times the shots of AD, less than the ratios 7.8 of the trained and
16.8 of the untrained scales but in the same direction, and AD is more
accurate than the Pauli twirl by 0.026, CI $[+0.017,+0.035]$. This difference
survives the standardized readout, so it is not one of scale; AD also fits the
training set better (0.921 against 0.908). At this depth the noiseless circuit
generalizes less well within the 100 training steps (training accuracy 0.910),
and AD exceeds it by 0.035, CI $[+0.024,+0.047]$, while the twirls do not
differ from it resolvably. A third of that difference is a better fit and two
thirds a smaller generalization gap (0.054 against 0.079), with a lower input
sensitivity ($\chi=1.28$ against 2.14). AD does not exceed the best noiseless
circuit (0.869 at $L=2$ to 4), and the gain needs many shots: evaluated with
$10^3$ shots per image, AD reaches 0.69 against 0.83 for the noiseless
circuit, and it is ahead only at $10^4$ shots and above. We report it as an
observation at a fixed training budget and eight initializations; we have not
tested whether it persists with longer training or more data.

\begin{table*}[t]
\caption{Trained classifiers at weaker damping (MNIST, standardized readout,
exact expectation values, eight initializations): test accuracy, operational
shot cost relative to the noiseless circuit at the same depth, shot factor
$\kappa$, and the squared ratio of the AD and Pauli-twirl scales before
training, $(\sigma_{\AD}/\sigma_{\rm Pauli})^2$, from the exact average over
parameters (Sec.~\ref{sec:shots}). The row $p=0.3$ repeats
Secs.~\ref{sec:clf} and \ref{sec:shots}.}
\label{tab:pdep_trained}
\begin{ruledtabular}
\begin{tabular}{llccccccccccc}
 & & \multicolumn{4}{c}{test accuracy} & \multicolumn{3}{c}{shot cost / noiseless} & \multicolumn{3}{c}{$\kappa$} & before\\
$p$ & $L$ & None & AD & Pauli & Depol & AD & Pauli & Depol & AD & Pauli & Depol & training\\
\colrule
0.3 & 4 & 0.869 & 0.876 & 0.870 & 0.870 & 967 & $1.7\times10^4$ & $1.4\times10^4$ & 50 & $1.6\times10^4$ & $1.1\times10^4$ & $1.3\times10^3$\\
0.05 & 4 & 0.869 & 0.866 & 0.866 & 0.864 & 4.7 & 4.8 & 4.1 & 4.6 & 4.7 & 4.4 & 1.06\\
0.1 & 4 & 0.869 & 0.872 & 0.863 & 0.863 & 13.1 & 21.5 & 14.9 & 15.6 & 21.2 & 17.5 & 1.69\\
0.1 & 8 & 0.831 & 0.867 & 0.840 & 0.838 & 115 & 466 & 495 & 62 & 484 & 459 & 16.8\\
\end{tabular}
\end{ruledtabular}
\end{table*}

The rescaling is cheap compared with generic error mitigation. Probabilistic
error cancellation of the twirled channel needs a quasi-probability norm
$\Gamma_{\rm PEC}=1/c^2$ per insertion, 1.43 at $p=0.3$. Over the 36 insertions
of the $L=4$ classifier, the number of shots grows by
\begin{equation}
\Gamma_{\rm PEC}^{2\times36}=(1-p)^{-72}\approx10^{11},
\label{eq:pec}
\end{equation}
consistent with the exponential lower bounds on
mitigation~\cite{takagi2023universal,quek2024exponentially}. A rescaling
learned under shot noise costs far less: relative to the noiseless circuit, a
factor of 177 to 417 in shots for the unital channels and 7.5 for AD at $L=4$
and $N_s=10^3$. It restores only the scale of the few features the task uses,
not the noiseless state.

\subsection{Accuracy and input sensitivity}

With the output scale removed, AD ends slightly above the noiseless circuit at
three and four layers on MNIST and at four layers on Fashion-MNIST
(Sec.~\ref{sec:clf}). The Pauli twirl isolates the non-unital term. Against
it, AD is higher by 0.007 on MNIST at $L=4$, CI $[-0.001,+0.015]$, and by
0.013 on Fashion-MNIST, CI $[+0.003,+0.023]$. The gain is small; against the
Pauli twirl it is resolved only on Fashion-MNIST, and we do not see it at one
or two layers. It rests on eight initializations, and it is a result for
exact expectation values: trained and tested with $10^3$ shots per image, AD
is 0.7 to 1.5 points below the noiseless circuit at $L=1$ to 4
(Sec.~\ref{sec:shots}). At
$p=0.1$ and $L=8$ the gain over the noiseless circuit is larger, 0.035, where
the noiseless circuit generalizes less well (Sec.~\ref{sec:shots}), and the
sensitivity $\chi$ is again lower under AD, 1.28 against 2.14 for the
noiseless circuit and 1.95 for the Pauli twirl. Gains from noise have been
reported for reservoir computers under amplitude
damping~\cite{domingo2023taking,monzani2024nonunital} and for the
generalization of variational
models~\cite{zhang2026finitenoise,zhu2026rethinking,scala2025equalization}; in
quantum kernel machines, decoherence acts as an implicit regularization,
although the test error there grows with the dephasing
rate~\cite{heyraud2022noisy}.

On MNIST the non-unital term also changes the input dependence of the trained
classifier. Figure~\ref{fig:clf}(c) shows the input sensitivity against
depth. For the noiseless circuit it grows from 0.65 at $L=1$ to 1.00 at
$L=4$, as each re-uploading layer adds to the dependence of the output on the
input. The unital channels follow this growth, reaching 0.94 at $L=4$. Under
AD the sensitivity barely grows, from 0.64 to 0.76. The difference from the
noiseless circuit is $-0.006$ at $L=1$, within the spread over
initializations, and grows to $-0.07$, $-0.16$, and $-0.25$ at $L=2$, 3, and 4,
each well outside it (Appendix~\ref{app:stats}). Against the Pauli twirl,
which isolates the non-unital term, the difference is $-0.03$, $-0.11$, and
$-0.18$ at $L=2$, 3, and 4, the latter two well resolved (paired $t$ test,
$P=0.007$ and $7\times10^{-4}$, unadjusted). For $L\ge2$
the twirls themselves lie slightly below the noiseless circuit, by at most
0.06, not resolved after correction for multiple tests (Appendix~\ref{app:stats}). The sensitivity also depends on
the gain of the readout, but the norm of the readout weights is similar for
all channels (4.4 to 4.7 at $L=4$), so the difference lies in the features.
The sensitivity of the standardized features themselves,
$\langle\lVert\partial\tilde{\mathbf z}/\partial\mathbf a\rVert\rangle$ with
$\tilde{\mathbf z}=\mathrm{diag}(\boldsymbol\sigma)^{-1}(\mathbf z-\boldsymbol\mu)$,
confirms this. At $L=4$ it is 2.79 for AD, against 4.13 for the noiseless
circuit and 3.98 for the Pauli twirl ($P=10^{-5}$ against the twirl, paired
$t$ test);
on Fashion-MNIST no channel differs from another.
The lower sensitivity makes the MNIST classifier more robust to a
miscalibrated encoding: a common offset of $\pm1$ on the four test angles at
$L=4$ costs AD 0.29 to 0.33 in accuracy, against 0.44 to 0.55 for the other
channels.
Depolarizing noise provably limits the effect of small adversarial
perturbations of the input~\cite{du2021quantum}. The offset here is large, and
at equal contraction only the non-unital channel reduces the drop.

On Fashion-MNIST at $L=4$, however, we find no such difference. The sensitivities $\chi$ are
2.56 for the noiseless circuit, 2.86 for AD, 2.89 for the Pauli twirl, and
3.16 for depolarizing noise, none resolved from the noiseless value, and AD is
not more robust to an offset. The effect on the input dependence is therefore
specific to the data. A candidate explanation is the effective shallowness
of noisy circuits~\cite{mele2026noise}. It holds for unital and non-unital
noise alike, but what survives the truncation differs: under the twirls the
features contain only contributions that survive the contraction of every
layer, whereas under AD they also carry the last-layer term of
Eq.~\eqref{eq:floor}, a first harmonic of a single angle, which dominates at
depth before training. We have not tested whether this term accounts for the
lower sensitivity on MNIST and its absence on Fashion-MNIST.

\section{Conclusion}
\label{sec:conclusion}

In this work, we have compared amplitude damping with its twirls, and in
particular with the Pauli twirl, which is damping at infinite temperature. The
zero-temperature bias acts on variational models mainly through the scale of
their features. For random parameters, the features under AD settle on a
floor, set at strong damping by the last layer and falling linearly in $1-2N$
with temperature at fixed $T_1$, while the twirls shrink every feature by a fixed factor per
layer, at every width up to eight qubits. A trainable output scale removes
most of the resulting differences in fit and accuracy in exact simulation, in
the single-qubit fits of Ref.~\cite{vanrossum2026biased} and in a four-qubit
classifier on MNIST and Fashion-MNIST; the rest reappears as a sampling cost.
Trained and tested with $10^3$ shots per image at $p=0.3$, the classifier
under AD stays within 1.5 percentage points of the noiseless accuracy at
depths 1 to 4, whereas the twirled classifiers lose up to 33 points at four
layers. The separation needs a depth
that grows roughly as $(np)^{-1}\ln(1/p)$, about 1500 layers for four qubits
at $p=10^{-3}$ and fewer at larger width, so it is a property of strong
damping or of very deep circuits.

Reversing the damping direction is a reparametrization whenever a Pauli frame
or spin time reversal can reach trainable gates and the readout; with a
symmetric initialization and an equivariant optimizer, trained models under
the two directions are then identically distributed, as in our classifier.
The direction matters only where the circuit traps the frame, as in the
eigensolver with the damping inside the decomposed two-qubit gate; there
reversed damping does worse also when the transverse field dominates, and on
an open chain the sign of the coupling is itself a gauge. Beyond the scale, the
bias leaves AD ahead of its twirl by up to three percentage points in exact
simulation, most at the separation depth among the depths we trained (2.6
points at $p=0.1$, $L=8$), and it lowers the input sensitivity on MNIST
only.

These results suggest three practices for comparisons of noise structure in
variational models. The first is to report the output scale, or to include a
trainable one, together with the shot budget it implies. The second is to
check whether the model class is invariant under the relevant frames before
attributing an effect to the direction of the noise. The third is to state
where the noise acts relative to the decomposition of the gates.

Several questions remain open. Our simulations are exact and small, with one
qubit for the fits, four for the classifier, and three for the eigensolver.
The width dependence is established before training, up to eight qubits;
training beyond four qubits was out of reach of our exact simulation, and our
shot model in training neglects the correlations between qubits.
The origin of the sensitivity effect on MNIST, and its absence on
Fashion-MNIST, remain unexplained; measuring the contribution of each layer to
the input dependence would be a first step. It also remains to be
seen whether the effect persists for more qubits and other data, whether
trained models follow the temperature dependence of the floor, and whether it
survives on hardware, where damping comes with dephasing. Finally, Refs.~\cite{tuysuz2024symmetry,ugail2026channel} studied
the opposite question to ours: whether noise breaks a symmetry that a model is
built to respect~\cite{meyer2023exploiting,larocca2022group}. The gauge freedom studied here is instead a symmetry of the
circuit that makes the direction of the noise invisible.

\begin{acknowledgments}

We thank Vietnam National University Ho Chi Minh City (VNU-HCM) for its support.
Claude Opus 5.5 (Anthropic) was used throughout the preparation of this work: to
search and survey the literature, and to
check its numerical claims and its characterizations of prior work against the
underlying data and sources.

\end{acknowledgments}

\section*{Data availability}
The code, the result files, and the scripts that reproduce every number and
figure in this paper are available at
\url{https://github.com/Identity-Operator/zero-vs-infinite-T-damping} and archived
on Zenodo at \href{https://doi.org/10.5281/zenodo.23053251}{doi:10.5281/zenodo.23053251}~\cite{nguyen2026zenodo}.

\appendix

\section{Reading of the protocol of Ref.~\cite{vanrossum2026biased}}
\label{app:vr}

We simulate the model of Ref.~\cite{vanrossum2026biased} (their App.~B1) in
the Bloch representation. Each of the $L+1=3$ trainable blocks
$\mathcal W_l=\mathcal U(W_l)$ of Eq.~\eqref{eq:vr_map} applies
\begin{equation}
W_l=R_Z(\theta_{3l})\,R_Y(\theta_{3l+1})\,R_Z(\theta_{3l+2}),\qquad l=0,1,2,
\label{eq:vr_block}
\end{equation}
and then the channel; after each of the first two blocks follow $R_X(x)$ and
the channel again. The output is $f(x)=r_z$. The channels act by
Eq.~\eqref{eq:bloch} with the entries of Table~\ref{tab:channels}.

Where the description in Ref.~\cite{vanrossum2026biased} leaves a choice, we
made the following one. The 250 inputs lie on a uniform grid over
$[-2\pi,2\pi]$. Each target $g'$ of Eq.~\eqref{eq:vr_target} has coefficients
drawn uniformly from $[-1,1]$ and is shifted to zero mid-range, normalized to unit
range, and multiplied by the range $R$,
\begin{equation}
g(x)=R\,\frac{g'(x)-\frac12\bigl(g'_{\max}+g'_{\min}\bigr)}{g'_{\max}-g'_{\min}},
\label{eq:vr_norm}
\end{equation}
where $g'_{\max}$ and $g'_{\min}$ are the extreme values on the grid. A step is one Adam
update on a batch of 25 inputs; the batches follow a seeded shuffle of the 250
inputs, renewed every ten steps and shared by all channels. We use their
settings for $L=2$ (learning rate 0.20, $\beta_1=0.45$, $\beta_2=0.95$, 150
steps). Initial parameters are drawn uniformly from $[0,2\pi)$, the same for
every channel. A target has converged at the first step at which the mean
squared error over all 250 inputs falls below threshold,
\begin{equation}
\frac1{250}\sum_{i=1}^{250}\bigl[f(x_i)-g(x_i)\bigr]^2<10^{-5} .
\label{eq:vr_conv}
\end{equation}
The largest
converging range is found by bisection on $[0,2]$ with ten halvings.

The raw model uses $f=\langle Z\rangle$ with the target centered on zero, or
$f=\langle Z\rangle+b$ with a trainable $b$, which is equivalent to the free
target offset $b$ of their Eq.~(7). The rescaled model is
Eq.~\eqref{eq:vr_scaled}, with $s$ and $b$ trainable from $s=1$ and $b=0$, and is
trained for up to 1000 steps at fixed $R$. The scale and offset are updated
with the same Adam settings as the circuit parameters.

\section{Proof of the frame gauge}
\label{app:gauge}

\begin{table*}[t]
\caption{Standardized-readout classifier on MNIST: difference from the
noiseless circuit in test accuracy ($\Delta$acc) and input sensitivity
($\Delta\chi$). $P$ values from paired $t$ tests over the eight
initializations, Holm-adjusted over the 22 tests of the table.}
\label{tab:stats}
\begin{ruledtabular}
\begin{tabular}{llcccc}
$L$ & channel & $\Delta$acc & 95\% CI & $P$ & $\Delta\chi$ ($P$)\\
\colrule
1 & AD & $+0.001$ & $[-0.003,+0.005]$ & 1 & $-0.006$ (1)\\
1 & Pauli & $-0.059$ & $[-0.178,+0.001]$ & 1 & $+0.058$ (1)\\
1 & Depol & $0$ & exact & -- & $0$ (--)\\
2 & AD & $+0.003$ & $[-0.003,+0.009]$ & 1 & $-0.072$ (0.004)\\
2 & Pauli & $+0.004$ & $[-0.003,+0.010]$ & 1 & $-0.043$ (0.26)\\
2 & Depol & $+0.000$ & $[-0.004,+0.004]$ & 1 & $-0.023$ (1)\\
3 & AD & $+0.009$ & $[+0.003,+0.015]$ & 0.04 & $-0.162$ (0.02)\\
3 & Pauli & $+0.001$ & $[-0.007,+0.008]$ & 1 & $-0.049$ (1)\\
3 & Depol & $+0.000$ & $[-0.007,+0.007]$ & 1 & $-0.037$ (1)\\
4 & AD & $+0.008$ & $[+0.002,+0.013]$ & 0.008 & $-0.247$ ($10^{-5}$)\\
4 & Pauli & $+0.001$ & $[-0.007,+0.009]$ & 1 & $-0.063$ (0.57)\\
4 & Depol & $+0.001$ & $[-0.006,+0.008]$ & 1 & $-0.062$ (1)\\
\end{tabular}
\end{ruledtabular}
\end{table*}

\paragraph{Setting.}
A model on $n$ qubits prepares $\rho_0=\ket{0}\bra{0}^{\otimes n}$, applies elements
$E_1,\dots,E_m$, and reads out
\begin{equation}
y_i=\Tr\bigl[O_i\,\Phi(\rho_0)\bigr],\qquad \Phi=E_m\circ\dots\circ E_1 ,
\label{eq:app_model}
\end{equation}
possibly followed by a learned affine map. Each $E_k$ is a fixed unitary $\mathcal U_G$, a data encoding
$\mathcal U_{S(x)}$ with $S(x)=\bigotimes_q R_{\mathbf n_q}(x_q)$, a trainable unitary from a
family $\mathcal W$ (general single-qubit blocks, or Pauli rotations $e^{-i\theta Q/2}$), or a
layer of the single-qubit channel $\AD$ on a set $S_k$ of qubits. The reversed model
replaces every noise layer by $\AD^{\rm flip}=\mathcal X\AD\mathcal X$.

\paragraph{Frames.}
Let $\mathcal P(\rho)=P\rho P$ for a Pauli string $P$, and $\mathcal K(\rho)=\bar\rho$. On Hermitian
operators $\mathcal K$ is real-linear, trace preserving and positive, $\mathcal K^2=\mathrm{id}$,
and for any channel $\mathcal E$ with Kraus operators $\{K_j\}$ one has
$\mathcal K\circ\mathcal E\circ\mathcal K=\bar{\mathcal E}$, the channel with Kraus operators
$\{\bar K_j\}$. A frame of unitary type is $g=\mathcal P$; of antiunitary type,
$g=\mathcal P\circ\mathcal K$. In both cases $g\circ g=\mathrm{id}$, and
\begin{equation}
g\circ\mathcal E\circ g'\colon\ \{K_j\}\mapsto\bigl\{P\,K_j^{(t)}\,P'\bigr\},
\label{eq:app_frame}
\end{equation}
with $K^{(t)}=K$ for the unitary type and $K^{(t)}=\bar K$ for the antiunitary type.

\paragraph{Telescoping.}
For any frames $g_0,\dots,g_m$ of one type,
\begin{equation}
\Phi^{\mathrm{rev}}=g_m\circ\big(g_m E^{\mathrm{rev}}_m g_{m-1}\big)\circ\dots\circ
\big(g_1 E^{\mathrm{rev}}_1 g_0\big)\circ g_0 ,
\label{eq:telescope}
\end{equation}
which is an identity since every inserted $g_k g_k$ is the identity.

\paragraph{The conditions.}
(i) Consider a noise layer on $S_k$ with $g_k=g_{k-1}$ acting as $X$ or $Y$ on each
$q\in S_k$. The Kraus operators of $\AD$ are real, and $Z\AD Z=\AD$, so
\begin{align}
X\AD X=Y\AD Y&=\AD^{\rm flip},
\label{eq:app_i}\\
g_k\circ\AD^{\rm flip}\circ g_{k-1}&=\AD
\label{eq:app_i2}
\end{align}
for both types. Off $S_k$ the frame passes unchanged.

(ii) For a fixed unitary $G$, $g_k\,\mathcal U_G\,g_{k-1}=\mathcal U_{P_kG^{(t)}P_{k-1}}$, which
equals $\mathcal U_G$ if and only if
\begin{equation}
P_k\propto G\,P_{k-1}\,G^{(t)\dagger} ;
\label{eq:app_ii}
\end{equation}
the frame propagates by conjugation and must remain a Pauli string. For Clifford $G$ this
holds for every $P_{k-1}$ of the unitary type; for real Clifford $G$ (CNOT) also of the
antiunitary type.

(iii) For an encoding, $g_k=g_{k-1}$ and $P_q R^{(t)}_{\mathbf n_q}(x)P_q=R_{\mathbf n_q}(x)$ for
all $x$. For the unitary type this requires $[P_q,\mathbf n_q\cdot\boldsymbol\sigma]=0$. For
the antiunitary type, $\overline{R_{\mathbf n}(x)}=\exp[+ix\,\overline{\mathbf n\cdot\boldsymbol\sigma}/2]$,
so the condition is
\begin{equation}
P_q\,\overline{\mathbf n_q\cdot\boldsymbol\sigma}\,P_q=-\mathbf n_q\cdot\boldsymbol\sigma ,
\label{eq:app_iii}
\end{equation}
which holds for $P_q=Y$ and every $\mathbf n_q$, since $\bar X=X$, $\bar Y=-Y$, $\bar Z=Z$,
$Y\sigma_{x,z}Y=-\sigma_{x,z}$, and $YYY=Y$.

(iv) A general single-qubit block $W$ is mapped to $P_{k,q}W^{(t)}P_{k-1,q}\in SU(2)$ for any
pair of frames: such blocks absorb, and may change, the frame on their qubits. A Pauli
rotation is mapped to
\begin{equation}
P\,e^{\mp i\theta Q^{(t)}/2}\,P=e^{-i\theta' Q/2},\qquad \theta'=\pm\theta ,
\label{eq:app_iv}
\end{equation}
if and only if $g_k=g_{k-1}$ and $PQ^{(t)}P=\pm Q$; it can carry a frame but not change it.

(v) By Eq.~\eqref{eq:telescope} the input is $g_0(\rho_0)=P_0\rho_0P_0$ (as $\bar\rho_0=\rho_0$).
If the first element on qubit $q$ is a general block $W$, the pair $W^{(t)}P_{0,q}$ is itself
a general block acting on $\ket 0$; if it is $R_X$ or $R_Y$, $X\ket0$ and $Y\ket0$ equal
$R_{X,Y}(\pi)\ket0$ up to phase and are absorbed by a shift of the angle; otherwise $P_{0,q}$
must fix $\ket0$, i.e. $P_{0,q}\in\{I,Z\}$. At the output, using
$\Tr[O\bar\tau]=\Tr[\bar O\tau]$ for Hermitian arguments,
\begin{equation}
\Tr\bigl[O\,g_m(\tau)\bigr]=\Tr\bigl[P_mO^{(t)}P_m\,\tau\bigr],
\label{eq:app_v}
\end{equation}
for any state $\tau$, which is $\eta\,\Tr[O\tau]$ if and only if $P_mO^{(t)}P_m=\eta O$.

\paragraph{Conclusion.}
If (i)--(v) hold, every bracket in Eq.~\eqref{eq:telescope} is an element of the original
model with reparametrized trainable parameters, the input frame is absorbed by the first
elements and the output frame contributes the sign $\eta$; hence
\begin{equation}
y^{\mathrm{rev}}_i(x;\theta)=\eta_i\,y_i(x;\theta')
\end{equation}
for all $x$, which is Eq.~\eqref{eq:reparam}. \hfill$\square$

\paragraph{Corollaries.}
The first statement of Corollary~\ref{cor:det} follows because $\theta\mapsto\theta'$ is a
bijection of the parameter space. For the second, write the map as $\theta'=S\theta+\pi\mathbf m$ with $S$
a diagonal sign matrix and $\mathbf m$ an integer vector, as for every frame of Table~\ref{tab:gauge}, together with
$W\mapsto W\,\mathrm{diag}(\boldsymbol\eta)$ for the readout. The training loss then satisfies
$\mathcal L^{\rm flip}(S\theta+\pi\mathbf m,\,W\mathrm{diag}(\boldsymbol\eta),\,b)=\mathcal L(\theta,W,b)$,
so its gradients transform with $S$ and $\mathrm{diag}(\boldsymbol\eta)$. An optimizer whose
update commutes with these maps (gradient descent and Adam without weight decay exactly, SPSA
in distribution) carries the trajectory from $(\theta_0,W_0,b_0)$ under $\AD$ onto the
trajectory from $(S\theta_0+\pi\mathbf m,\,W_0\mathrm{diag}(\boldsymbol\eta),\,b_0)$ under
$\AD^{\rm flip}$. If the map preserves the distribution of $(\theta_0,W_0,b_0)$, the trained
models are identically distributed. For Corollary~\ref{cor:rand}, write a qubit channel as in Eq.~\eqref{eq:bloch}.
Unitary Pauli conjugation and complex conjugation act on $(T,\mathbf t)$ as
\begin{align}
\mathcal P\colon&\ (T,\mathbf t)\mapsto(DTD,\,D\mathbf t),
\label{eq:app_D}\\
\mathcal K\colon&\ (T,\mathbf t)\mapsto(FTF,\,F\mathbf t),
\label{eq:app_F}
\end{align}
with $D$ a diagonal sign matrix and $F=\mathrm{diag}(1,-1,1)$; both preserve
$\lVert\mathbf t\rVert$. The Pauli twirl has $\mathbf t=0$, while
$\lVert\mathbf t\rVert=p$ for amplitude damping.

\paragraph{Time reversal and CNOT.}
By Eq.~\eqref{eq:theta_cnot}, time reversal leaves a local Pauli residue on the
input side of each CNOT. A $Z$ residue
passes through amplitude damping ($Z\AD Z=\AD$); an $X$ residue does not. The residue must
therefore reach a general trainable block without crossing a noise location, which is the
case in our classifier (the trainable layer immediately precedes the ring) and fails when
noise is inserted after each CNOT (Table~\ref{tab:gauge}).

\paragraph{Scope.}
The proposition is a sufficient condition. The rows marked ``none'' in
Table~\ref{tab:gauge} establish that no frame of either type exists; they do not by
themselves prove that the classes differ. For the single-qubit regression model the
output range proves it. For the eigensolver at $\gamma=0.3$, with L-BFGS-B
from 12 random initializations and the channel applied once after each
$R_{ZZ}$ or after each CNOT, the lowest energies are equal where the gauge
holds ($-3.1770$ for both) and differ by 1.98 where it does not ($-3.087$
against $-1.109$), reproducing the direction of the finding of
Ref.~\cite{vanrossum2026biased}. For the classifier with noise inside the
ring, Sec.~\ref{sec:gauge_clf} measures the difference directly.

\section{Statistical tests}
\label{app:stats}

We collect here the hypothesis tests behind the comparisons in the main text.
To avoid confusion with the noise strength $p$, we denote $P$ values by a
capital $P$. For the classifier, accuracy differences come with the 95\% CI of
the two-level bootstrap described in Sec.~\ref{sec:setting} ($2\times10^4$
resamples) and the $P$ value of the paired $t$ test across the eight
initializations, which the conditions share; sensitivity differences come with
the paired $t$ test. The $P$ values of Table~\ref{tab:stats} are Holm-adjusted
over its 22 tests. Five remain below 0.05: the accuracy of AD at $L=3$ and 4,
and its sensitivity at $L=2$, 3, and 4.

On Fashion-MNIST at $L=4$, the differences from the noiseless circuit are
$+0.017$, CI $[+0.007,+0.029]$, $P=0.03$, for AD; $+0.005$, CI
$[-0.004,+0.014]$, $P=0.28$, for the Pauli twirl; and $+0.009$, CI
$[-0.0001,+0.018]$, $P=0.08$, for depolarizing noise (paired $t$,
Holm-adjusted over the three channels). The sensitivity differences are
$+0.30$ ($P=0.25$), $+0.33$ ($P=0.21$), and $+0.60$ ($P=0.21$). AD minus the Pauli twirl in accuracy is $-0.001$, CI
$[-0.007,+0.005]$, at $L=2$; $+0.008$, CI $[-0.0002,+0.017]$, at $L=3$;
$+0.007$, CI $[-0.001,+0.015]$, at $L=4$ on MNIST; and $+0.013$, CI
$[+0.003,+0.023]$, on Fashion-MNIST. At $L=1$ on MNIST it is dominated by the
diverged Pauli run.

For the classifier with the noise after each CNOT of the ring (Sec.~\ref{sec:gauge_clf}),
AD$^{\rm flip}$ minus AD is $-0.006$ in accuracy, CI $[-0.013,+0.001]$,
$P=0.10$, and $-0.008$ in sensitivity, $P=0.71$. With the standard placement
and the raw readout, AD$^{\rm flip}$ minus AD is $-0.003$, CI
$[-0.008,+0.002]$, $P=0.21$.

For the single-qubit fits, convergence of AD and of the Pauli twirl on the
same 200 targets is compared with McNemar's exact test. Within 1000 steps, no
difference is significant at either range ($P\ge0.34$). Within 150 steps, the
only significant difference is $-13.5$ percentage points at $p=0.2$ and
$R=0.468$ ($P=0.0013$); the others have $P\ge0.077$. For the raw model, the
largest converging ranges of AD and the Pauli twirl are compared with the
paired Wilcoxon test: with a free target offset, $P=0.71$, 0.049, and
$6\times10^{-14}$ at $p=0.05$, 0.1, and 0.2; with a centered target,
$P=0.19$, $9\times10^{-7}$, and 0.04.

For the eigensolver, the relative errors of reversed AD and AD over the 50
initializations are compared with the Mann--Whitney test. With the noise
inside $R_{ZZ}$, $P<10^{-16}$ at every $\gamma$; with the noise after it,
$P\ge0.69$.

\section{Feature scale before training}
\label{app:width}

The lines in Fig.~\ref{fig:width} are exact averages, over parameters drawn
uniformly from $[0,2\pi)$, of
\begin{equation}
\sigma^2=\mathbb E_\theta\,\frac1n\sum_{j=1}^n\mathrm{Var}_{\mathbf a}\,z_j ,
\label{eq:sigma_rms}
\end{equation}
where the variance runs over the first 300 training images. We compute them in
the Heisenberg picture from the second moments of the Pauli-string
coefficients of $Z_j$, with the parameters shared between two images. Every
uniformly random $R_ZR_YR_Z$ block removes the cross moments between different
strings because, averaged over its angles, the entries $R_{bc}$ of its
adjoint action have vanishing first moments and different columns are
uncorrelated, $\mathbb E_\theta[R_{xb}R_{yc}]=0$ for $b\neq c$. Between two
blocks the circuit acts site by site (channel, $R_X(a_j)$, channel), so the
second moments of two images propagate through the elementwise product of the
two composite site maps, followed by the entangling ring, which permutes the
strings up to a sign. The second moments therefore obey a closed recursion, which is exact up
to the sampling of image pairs: all pairs for $n\le5$, and 5000 random pairs
for $n\ge6$. Against brute-force averages of the density-matrix simulation,
the recursion agrees within one to two standard errors in all 18 cells at
$n=2$ and 3 and $L=1$ to 3. An independent density-matrix simulation at $n=4$, with 200 parameter draws, agrees within two standard errors in all 12 cells for the noiseless circuit, AD, and the Pauli twirl at $L=1$ to 4, and in all four cells for AD and the Pauli twirl at $p=0.1$ and $L=4$ and 8.

Under AD the adjoint channel maps $Z$ to $c^2Z+pI$. The next random block
averages the cross moments between the $I$ and $Z$ branches to zero, so that
$\sigma_{\AD}^2=\sigma_{\rm Pauli}^2+\sigma_{\rm br}^2$, where $\sigma_{\rm br}$
collects the paths through at least one branching into the identity. The
branching after the last entangler gives Eq.~\eqref{eq:floor}. Averaged over
the final block, with $\mathbb E\,r_z^2=\frac12$, $\mathbb E\,r_y^2=\frac14$, and
$\mathbb E\,r_zr_y=0$, it yields
\begin{multline}
\sigma_\infty^2=\frac{p^2(1-p)^2}{n}\sum_j\Bigl[\frac{(1-p)^2}{2}\,
\mathrm{Var}(\cos a_j)\\
+\frac{1-p}{4}\,\mathrm{Var}(\sin a_j)\Bigr],
\label{eq:floor_var}
\end{multline}
which at $p=0.3$ is the floor of $\sigma_{\AD}$ within 0.5\% for $n\ge4$ (with all image pairs);
branchings in earlier layers are attenuated by the rest of the circuit. At
weaker damping they are not, and they add the term $b_{\rm br}\,p$ of
Sec.~\ref{sec:shots}. For uniformly
distributed angles, $\sigma_\infty=0.096$. The measured points in
Fig.~\ref{fig:width} use five parameter draws, the same parameter vectors at
every depth (those of depth $L$ are the leading entries of those of depth 6),
so their deviations are correlated across $L$. The distribution over the
parameters is heavy-tailed, and five draws tend to fall below the average, as
for the Pauli twirl at $n=8$ and $L\ge4$. An independent Pauli-path simulation
reproduces these five draws to $2\times10^{-12}$ and, over 120 further draws
at $L=4$, agrees with the exact average within one standard error.

\bibliography{refs}

\end{document}